\documentclass[useAMS,usenatbib]{mn2e}
\usepackage{graphicx,color}
\usepackage{amsmath}
\usepackage{amssymb}
\usepackage{comment}
\usepackage{subfigure}
\usepackage{caption}

\include{defns}
\def\gsim{\;\rlap{\lower 2.5pt
 \hbox{$\sim$}}\raise 1.5pt\hbox{$>$}\;}
\def\lsim{\;\rlap{\lower 2.5pt
   \hbox{$\sim$}}\raise 1.5pt\hbox{$<$}\;}
\defcitealias{liu09a}{Liu et al. (2009a)}
\defcitealias{liu09b}{Liu et al (2009b)}

\title[The Distribution of Bubble Sizes During Reionization]{The Distribution of Bubble Sizes During Reionization}
\author[Lin, Oh, Furlanetto \& Sutter]{Yin Lin$^{1,2}$, S. Peng Oh$^{1}$, Steven R. Furlanetto$^{3}$, P.M. Sutter$^{4,5,6}$ \\
$^{1}$Department of Physics; University of California; Santa Barbara, CA 93106, USA.\\
$^{2}$ Department of Physics; University of Chicago, Chicago, IL 60637, USA.\\
$^{3}$ University of California Los Angeles, Department of Physics and Astronomy, Los Angeles, CA 90095, USA.\\
$^{4}$ INFN - National Institute for Nuclear Physics, via Valerio 2,
I-34127 Trieste, Italy \\
$^{5}$ INAF - Osservatorio Astronomico di Trieste, via Tiepolo 11,
1-34143 Trieste, Italy \\
$^{6}$ Center for Cosmology and Astro-Particle Physics, Ohio State
University, Columbus, OH 43210}

\begin{document}
\bibliographystyle{mn2e}

\pagerange{000--000} \pubyear{0000}
\maketitle

\label{firstpage}

\begin{abstract}
A key physical quantity during reionization is the size of HII regions. Previous studies found a characteristic bubble size which increases rapidly during reionization, with apparent agreement between simulations and analytic excursion set theory. Using four different methods, we critically examine this claim. In particular, we introduce the use of the watershed algorithm -- widely used for void finding in galaxy surveys -- which we show to be an unbiased method with the lowest dispersion and best performance on Monte-Carlo realizations of a known bubble size PDF. We find that a friends-of-friends algorithm declares most of the ionized volume to be occupied by a network of volume-filling regions connected by narrow tunnels. For methods tuned to detect the volume-filling regions, previous apparent agreement between simulations and theory is spurious, and due to a failure to correctly account for the window function of measurement schemes. The discrepancy is already obvious from visual inspection. Instead, HII regions in simulations are significantly larger (by factors of $10-1000$ in volume) than analytic predictions. The size PDF is narrower, and evolves more slowly with time, than predicted. It becomes more sharply peaked as reionization progresses. These effects are likely caused by bubble mergers, which are inadequately modeled by analytic theory. Our results have important consequences for high-redshift 21cm observations, the mean free path of ionizing photons, and the visibility of Ly$\alpha$ emitters, and point to a fundamental failure in our understanding of the characteristic scales of the reionization process.
\end{abstract}


\section{Introduction}
\label{section:intro}

Reionization is the last global event in the history of our universe, akin to primordial nucleosynthesis or recombination, in which virtually all baryons participated. What would be at the top of our wish list in understanding this still mysterious epoch? Clear frontrunners are: (i) the progress of reionization, or $Q_{\rm HII}(z)$, the growth of the HII filling fraction with time. Its time derivative is closely related to the comoving emissivity, and functions as reionization's ``Madau plot". (ii) The topology of reionization, or the distribution of HII bubble sizes. This is reionization's ``mass function", and illuminates many properties of the underlying galaxy distribution and radiative transfer during reionization. Current influential models suggest that reionization proceeds outward from overdense regions (``inside out"); galaxy clustering produces large ($\sim 10$s Mpc comoving) HII regions whose characteristic size increases as reionization progresses. The distribution of bubble sizes was first predicted analytically via the excursion set formalism (\citet{furl04-bub}; hereafter `FZH04'), and has since been largely corroborated by comparisons with semi-numeric and radiative transfer simulations of reionization \citep{mesinger07,mcquinn07,zahn07,zahn11}. These all paint an apparently consistent picture of an approximately lognormal bubble size distribution which peaks at a characteristic scale, and becomes increasingly peaked as reionization progresses. Indeed, given the complexity of the reionization process, the agreement between simulations and analytic theory (which for instance assumes spherical bubbles, as opposed to the complex, non-spherical shapes seen in simulations) is remarkable. Interestingly, a dissenting view came from \citet{iliev06} and \citet{friedrich11}, which found from friends-of-friends analysis of large scale radiative transfer simulations two populations of HII regions: numerous, mid-sized ($\sim 10$ Mpc) regions, and rare, very large regions several tens of Mpc in size, which contained a considerable fraction of the volume. Recently, \citet{paranjape14} presented a modified excursion set theory calculation which predicts typical bubble sizes more than a factor two larger than earlier calculations-- a trend which is borne out and amplified in this work.

The characteristic scale of bubbles is of considerable importance: the characteristic scale of bubbles affects the amplitude of 21cm brightness temperature fluctuations and the scale at which they peak. The size of bubbles directly affects our ability to do direct imaging and tomography, as well as our ability to cross-correlate the fluctuating 21cm signal against the galaxy population (particularly against lines whose visibility is directly affected by bubble size, for instance Ly$\alpha$). Large bubbles may also assist with the recovery of large-scale 21cm modes which are otherwise inaccessible due to foreground subtraction \citep{petrovic11}. Thus, theoretical calculations of bubble size have been crucial in setting the science agenda for low-frequency interferometers, and indeed in instrument design itself. It is one of the most important theoretical inputs to instrumentalists. 

In this paper, we provide a critical re-analysis of some of the methods previously used to characterize bubble size in numerical simulations, and in addition employ a new algorithm, the watershed method, which has previously been used to detect voids in galaxy surveys \citep{platen07,neyrinck08,sutter15}. The definition of bubble size is somewhat ambiguous: different methods weight bubble topology in different ways, and will return different answers for the bubble size. For instance, if connectedness is our only metric, then past a critical point, virtually all of the ionized volume is consumed by a single large bubble, a result which can be understood from percolation theory \citep{furlanetto16}. Nonetheless, for metrics which are most physically and observationally relevant, we find effective bubble radii larger than that predicted by excursion set theory by a factor of a few, up to an order of magnitude (and thus larger bubble volumes by 1-3 orders of magnitude). The outline of this paper is as follows: in \S\ref{section:method}, we discuss the four methods we use to characterize bubble sizes, and how they are tested and calibrated against control samples. In \S\ref{section:results}, we then apply these methods to simulations of reionization. In \S\ref{section:convergence}, we describe convergence studies. In \S\ref{section:discussion}, we discuss implications of our findings, and conclude in \S\ref{section:conclusions}. Unless otherwise specified, all distances are in comoving units.

\section{Method}
\label{section:method} 

\begin{figure}
\begin{center}
\includegraphics[width=0.5\textwidth]{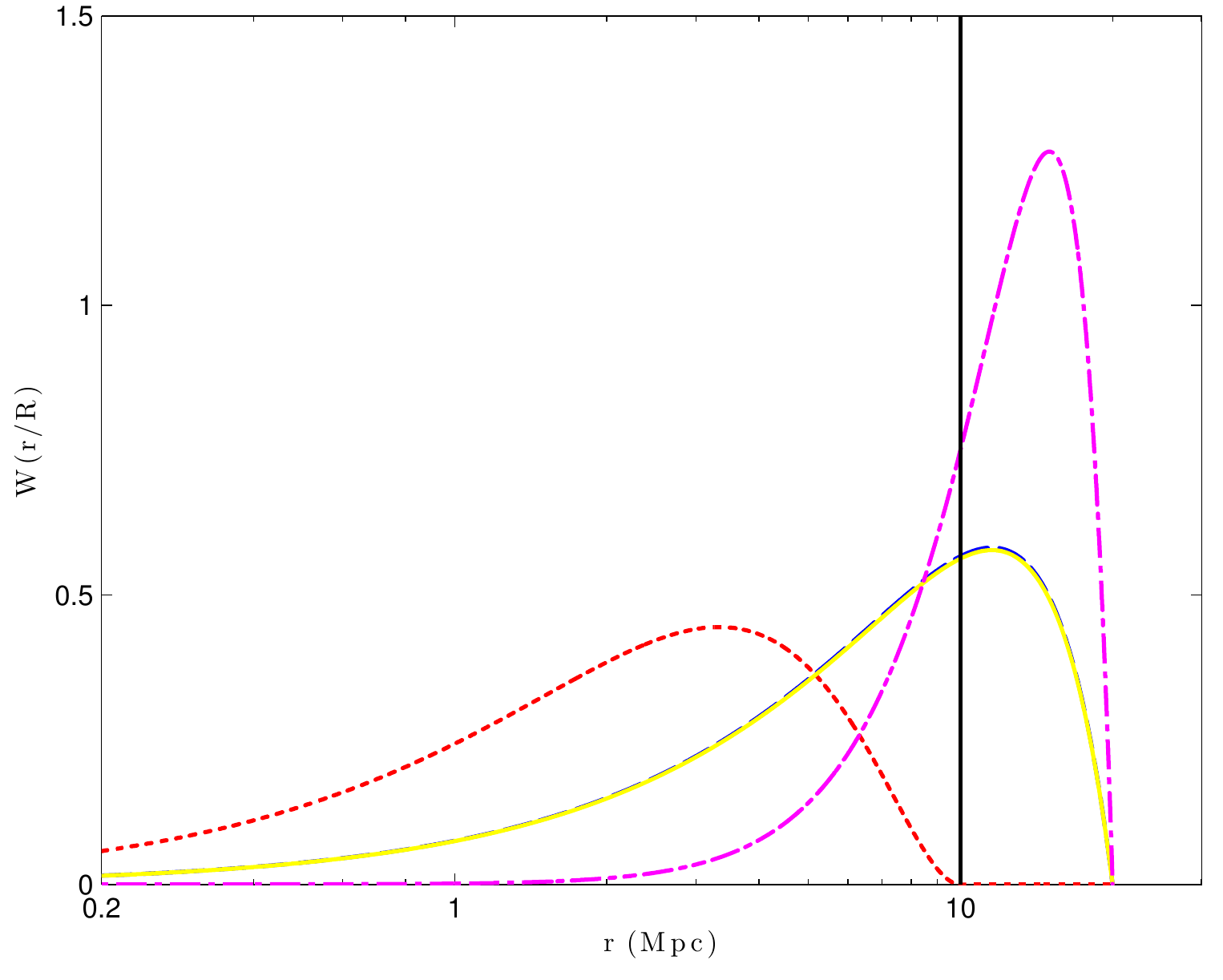}
    \caption{The bubble PDF for numerical calculations of the mean free path (MFP) and distance transform (DT) of a sphere of radius $R=10$ Mpc. The red dotted line is for the DT, the blue dashed line and the yellow solid line are for numerical and analytic calculations of the MFP respectively (note their excellent agreement), and the pink dash-dot line is the MFP window function used by \citet{mesinger07}. The watershed and FoF algorithms give delta functions at the true bubble size.}
    \label{fig:window_functions}
\end{center}
\end{figure}

\begin{figure}
\begin{center}
\includegraphics[width=0.5\textwidth]{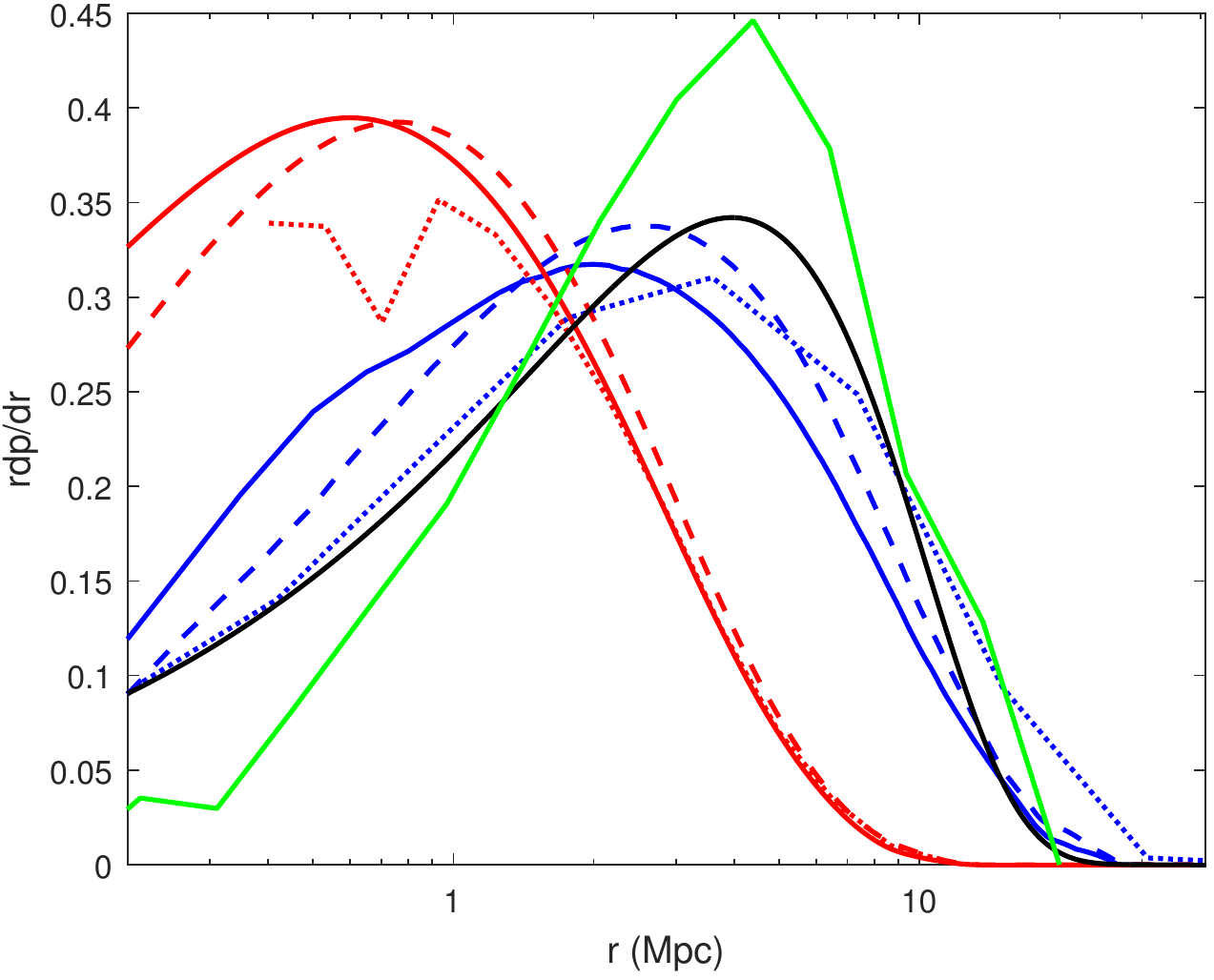} \\
\caption{Bubble PDFs for Monte Carlo toy box sampled at $Q_{\text{HII}} = 0.5$ (z=11 for $\zeta = 40$), with an Lagrangian FZH04 bubble mass function. Each line is normalized to be integrated to unity. The black line is the unconvolved excursion set density function. The green solid line is the watershed PDF (smoothing parameter $h=0.7$). The blue and red solid lines are the convolution of direct sampling of the excursion set prediction with the MFP and DT window functions respectively. The blue and red dashed lines are the convolution of the watershed PDF with the MFP and DT window functions. The blue and red dotted lines are the results of MFP sampling with $10^7$ samples and DT. The FoF algorithm identifies most of the volume in a single large bubble and is not shown.}
\label{fig:mfp_monte}
\end{center}
\end{figure}

We use 4 methods of analyzing reionization simulations to obtain the bubble size distribution. Two, the distance transform (DT; \citealt{zahn07,zahn11}) and the mean free path (MFP; \citealt{mesinger07}) methods have previously been used to obtain singly peaked distributions with characteristic bubble sizes in good agreement with excursion set theory. The third, friends of friends (FoF; \citealt{iliev06,friedrich11}) was shown to yield a bimodal distribution. The last, the watershed algorithm, has never previously been used in this context. In this section, we compare and contrast these different algorithms by: (i) deriving their window function $W(r/R)$, i.e. the derived bubble size distribution for a single spherical bubble of radius $R$; this determines how accurately the underlying bubble size distribution can be determined. (ii) validating their performance on a Monte-Carlo realization of spherical bubbles drawn from the distribution predicted by excursion set theory. 
At this stage, we deliberately restrict ourselves to spherical bubbles since sphericity is an explicit assumption of excursion set theory. If an algorithm already fails in this idealized regime, there is no reason to trust its results in more complicated settings. Note that the DT and MFP methods do not yield a unique segmentation of the ionization field into bubbles of different sizes (membership in a bubble of size R is only assigned one voxel at a time), while the FoF and watershed methods {\it do} result in a unique partitioning which can be visualized. 

We focus on the statistic $dP/dR$, i.e. the probability that an ionized voxel selected at random lies in a bubble with radius in the range $(R,R+dR)$. We generally plot $dP/d(\ln R)$, normalized to unity, which is equivalent to the fraction of volume occupied by bubbles with these radii. Given an intrinsic PDF $[dP/d(\ln R)]_{\rm i}$, the recovered bubble PDF is: 
\begin{equation}
\left(\frac{dP}{d{\ln R}} \right)_{\rm obs} = \int d({\ln r}) \left(\frac{dP}{d{\ln r}} \right)_{\rm i} W({r}/{R}) 
\end{equation}
where $W(r/R)$ is the window function (i.e., $(dP/d(\ln R))_{\rm obs}$  for a single bubble of radius R). For a spherical bubble distribution with a differential number density as a function of radius $d{\rm n}/d{\rm r}$, 
\begin{equation}\label{eqn:dpdr}
\left(\frac{dP}{d{\ln r}} \right)_{\rm i} = \frac{V}{Q_{\rm HII}} \frac{d {\rm n}}{d {\ln r}},
\end{equation}
where $V=4/3 \pi r^{3}$, and $0 < Q_{\rm H II} < 1$ is the volume filling fraction of ionized regions.  

Note that the number density $dn/dr$ must be in Eulerian coordinates to compare to results from our simulations. However, the analytic FZH04 model predictions are in Lagrangian coordinates. We use the following procedure to transform between the two. 
Let $r_0$ be the comoving Lagrangian radius and $r$ be the comoving Eulerian radius.

From the standard spherical collapse model we obtain \citep{paranjape14}:
\begin{equation}
\frac{r}{r_0} = 2\cdot 6^{2/3}\frac{(\theta - \sin\theta)^{2/3}}{1-\cos\theta}
\end{equation}
where $\theta$ is the development angle which, at a given redshift $z$ and linear density contrast $\delta_0$, can be determined from
\begin{equation}
\frac{1}{1+z} = \frac{3\cdot6^{2/3}}{20\delta_0}(\theta - \sin\theta)^{2/3}
\end{equation}
where we use $\delta_0 = D(z)B(m,z)$, where $D(z)$ is the growth factor and $B(m,z)$ is the linear fit to barrier as presented in FZH04. We can rewrite equation (\ref{eqn:dpdr}) as
\begin{equation}
\left(\frac{dP}{d{\ln r}} \right)_{\rm i} = \frac{V}{Q_{\rm HII}} \frac{d {n}}{d {\ln r}} =  \frac{V_0}{Q_{\rm HII}}\frac{dn}{d\ln r_0}\left(\frac{V}{V_0}\right)\left(\frac{d\ln r}{d\ln r_0}\right)^{-1}
\end{equation}
where $dn/dr_0$ is the Lagrangian number density in FZH04, $V_0 = (4/3)\pi r_0^3$ is the Lagrangian volume, and the Jacobian $d \ln r / d\ln r_0$ is equal to
\begin{equation}
\frac{\ln r}{\ln r_0} = 1 - \left|\frac{d\sigma^2}{dr_0}\right|\frac{dB}{d\ln \sigma^2}\left(1-\frac{3}{2}\frac{\theta(\theta-\sin\theta)}{(1-\cos\theta)^2}\right),
\end{equation}
and $\sigma(m)$ is the density fluctuation variance linearly extrapolated to $z=0$. Finally, the filling fraction\footnote{For subtle reasons, equation (\ref{eqn:Q_HII}) and the frequently used alternative $Q_{\rm HII}=\zeta f_{\rm coll}$, where $\zeta$ is the ionizing efficiency and $f_{\rm coll}$ is the collapse fraction of dark matter halos above a threhold mass, which in principle should be equivalent, can differ slightly \citep{paranjape16}.} is given by 
\begin{equation}
Q_{\rm HII} = \int^\infty_{r_{\rm min}} V\frac{dn}{d\ln r} (d\ln r) 
\label{eqn:Q_HII}
\end{equation}
where $r_{\rm min}$ corresponds to the smallest possible ionized bubble, which is ionized by a single halo with virial temperature $T_{\rm vir} = 10^{4}$K.

The transformation between Lagrangian and Eulerian coordinates is a small ($< 10\%$) correction. It is only used when comparing FZH04 to simulation results (and not, for instance, when creating a Monte-Carlo catalog from a given bubble PDF, e.g. \S\ref{section:monte_carlo}).

\subsection{Distance Transform (DT)} 
\label{section:spa} 

The use of spherical averaging (\citealt{mcquinn07,zahn07}) was motivated by how bubbles are defined and generated within the excursion set formalism, and in principle should provide the closest correspondence to bubble sizes predicted by excursion set theory. At each point in the simulation box, one `draws' progressively smaller spheres of radius R, and smooths the ionization field within the sphere. The voxel is deemed to belong to the {\it largest} sphere of radius R where the smoothed ionization fraction exceeds a threshold value $x_{\rm HII}^{t}$. The short mean free path of UV ionizing photons at ionization fronts means that the IGM is predominantly two-phase, and values of $x_{\rm HII}^{t}$ between 0 and 1 are generally finite resolution effects. For simplicity, we use $x_{\rm HII}^{t} = 1$ as it gives an analytic window function; lower values of $x_{\rm HII}^{t}$ give similar results \citep{zahn11}. For this threshold value, spherical averaging is exactly the same as the distance transform later used as a step in the watershed algorithm; each voxel is labelled by the shortest distance to a neutral voxel. We henceforth refer to this as the distance transform (`DT') method. 

Consider a point in a sphere of radius R, which is a distance $r$ away from the closest boundary of the sphere, and a distance $(R-r)$ from the sphere's center. The probability of hitting such a point at random is $dP=4 \pi (R-r)^{2} dr/(4/3) \pi R^{3}$, or \citep{friedrich11}: 
\begin{equation}
W\left(\frac{r}{R}\right)= \left( \frac{d\, {\rm P}}{d \, {\rm ln \, r}} \right) = 3 \left( \frac{r}{R} \right) \left( 1- \frac{r}{R} \right)^{2}.
\label{eqn:window_func_SPA}    
\end{equation}
In Fig \ref{fig:window_functions}, we show a numerical calculation of $W(r/R)$, which agrees with equation (\ref{eqn:window_func_SPA}). Clearly the DT is a poor method for estimating bubble sizes: it has a broad smoothing kernel spanning $\sim 2$ dex, and peaks at $r=R/3$. This distortion arises because there is more volume and hence more voxels at large radii, where the distance to the HII boundary is smaller. Since the true PDF is convolved with this window function, the observed PDF will be smeared out and biased toward lower bubble sizes. This is clearly apparent in Fig. \ref{fig:mfp_monte}, where the DT method does not recover the underlying bubble distribution of a Monte-Carlo catalog where bubbles are randomly drawn from the FHZ04 PDF (see \S\ref{section:monte_carlo} for more details), but a distorted version heavily biased toward small bubble sizes.  Indeed, we see that if we take the true bubble distribution and convolve it with the window function (equation \ref{eqn:window_func_SPA}), then this matches our numerical result very well. We regard the DT as an inferior technique for inferring bubble sizes. 

\subsection{Mean Free Path (MFP)}
\label{section:mfp} 

This method, first used by \citet{mesinger07}, consists of selecting a random ionized point, choosing a random direction, and finding the distance along that direction to the boundary of the HII region. Repeatedly performing this in a Monte-Carlo fashion allows us to find the PDF of `mean free paths', which acts as a proxy for bubble size. This measure is attractive because photon mean free path (which, until the late stages of reionization, is primarily determined by bubble size rather than absorption by Lyman limit systems) is an important physical variable. It is crucial in setting the strength of the ionizing background $\Gamma \propto \epsilon \lambda$ (where $\epsilon$ is the emissivity and $\lambda$ is the mean free path). 

We present an analytic derivation of the window function in Appendix \ref{appendix:mfp_window_func}. In Fig \ref{fig:window_functions}, we compare this expression with direct numerical calculations of the MFP PDF for a single spherical bubble; the two agree extremely well. Note that our window function differs from the erroneous expression derived by \citet{mesinger07}, which disagrees with direct numerical calculations. 
The \citet{mesinger07} window function is somewhat narrower and peaked toward larger bubble sizes (see Fig \ref{fig:window_functions}); because this assumed window function is convolved with the excursion set theory prediction, it skews predicted bubble sizes to be larger, somewhat reducing the discrepancy between theory and simulations.\footnote{\citet{mesinger07} did, however, use the MFP method to observe the main effects we see (\S\ref{section:real_sims}): larger bubble sizes in the simulations, with slower evolution in the bubble PDF compared to theory.} 
In Fig \ref{fig:window_functions}, we see that the MFP is unbiased and peaks at the correct bubble size. However, it is also asymmetric and comparable in width ($\sim$ 2 dex in $r/R$) to the DT method. The large number of voxels close to the bubble boundary have opposing short and long lines of sight which give the correct mean, but an asymmetric distribution. The MFP method thus introduces some smoothing and distortion of the underlying PDF, but for our purposes it should be acceptable. In Fig \ref{fig:mfp_monte}, we see that the MFP method reasonably approximates the input PDF in a Monte-Carlo simulation. Also, the input PDF convolved with our window function matches our numerical result, confirming that we have the correct window function. 

Note that \citet{friedrich11} find that the bubble PDF derived from the MFP method is sensitive to the threshold $x_{\rm th}$ used to demarcate the transition between neutral and ionized regions. This is largely a resolution effect: as resolution increases, the fraction of partially ionized voxels decreases. Since the semi-numeric simulations we use by their nature divide the IGM into fully ionized or fully neutral regions, with only a negligible fraction of partially ionized voxels, we find that this is much less of a problem in our work. 

\subsection{Friends of Friends (FoF)} 
\label{section:fof} 

The friends of friends algorithm groups together neighboring cells of the same equivalence class. It has been widely used for halo-finding in cosmological N-body simulations, where all particles separated by less than a given linking length $b \bar{l}$ (where $\bar{l}=\bar{n}^{-1/3}$ is the mean inter-particle separation and $b$ is a free parameter) are assigned to the same halo. Here, it links together neighboring cells which are either ionized or neutral. Unlike other methods, FoF is relatively insensitive to the topology of ionized regions. A network of connected tunnels will be classified as a single monolithic entity with a large volume, whereas the other methods we use  will consider them to be a large collection of smaller ionized regions. Using FoF, \citet{iliev06} and \citet{friedrich11} found a surprising result: the bubble size distribution is {\it bimodal}, with the vast majority of the ionized volume in a single large bubble as reionization progresses. This is in sharp contrast to the findings of other methods, and predictions from the excursion set formalism, and raises the question of whether some crucial aspect of reionization has been missed.

\begin{figure}
\begin{center}
\includegraphics[width=0.5\textwidth]{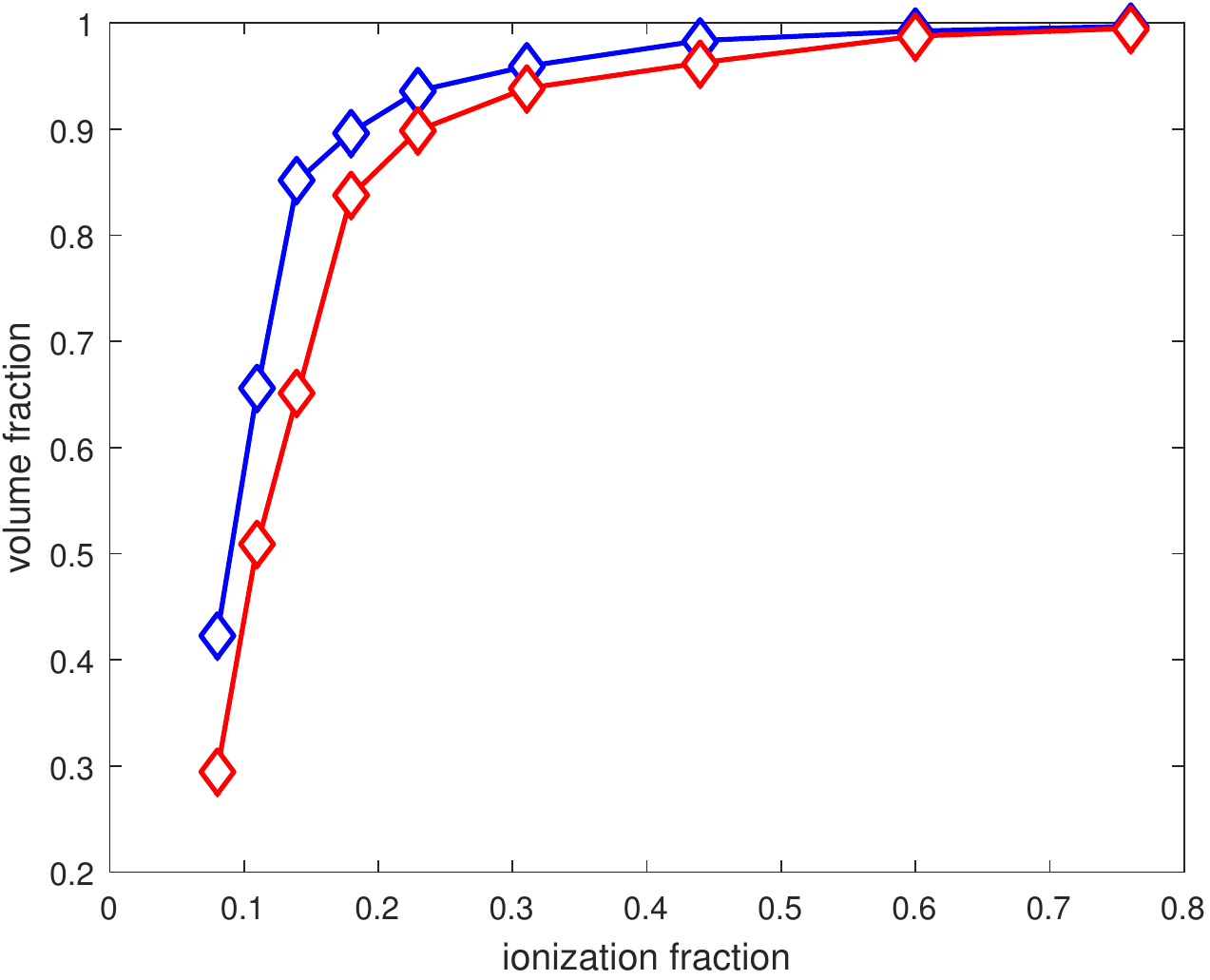}
    \caption{The fractional volume of ionized voxels inside the largest ionized bubble as defined by FoF method. The blue plot is for 26 connected neighborhood voxels (i.e., each voxel must have 26 neighbors, consisting of face, edge, or vertex adjacent voxels) whereas the red plot is for 6 connected neighborhood voxels (requiring only face adjacent voxels).}
    \label{fig:FoF}
\end{center}
\end{figure}

For a single sphere, FoF correctly segments the entire bubble; its window function is a delta function. For the Monte-Carlo catalog, the distribution is bimodal and dominated by a single large bubble; past an ionization fraction of $\sim 30\%$, it occupies virtually all of the volume. Fig \ref{fig:FoF} shows the volume fraction of the largest ionized bubble for realistic semi-numeric simulations of reionization; it starts to dominate and quickly rise toward unity once the volume fraction exceeds $\sim 15\%$. If connectedness is our only criterion, then throughout reionization, most of the volume is in a single bubble. Thus, in Fig \ref{fig:mfp_monte}, the bubble size distribution is essentially a delta function with a bubble volume $x_{\rm HII} V_{\rm box}$. This structure is created by bubble overlap (even in the absence of clustering) and regardless of the details of the underlying PDF. For instance, it arises even if all bubbles are identical (so the PDF is a delta function). It is also robust to the details of the FoF algorithm (for instance, in Fig. \ref{fig:FoF}, we show find fairly similar behavior given two different criteria for connectedness, requiring either 6 or 26 nearest neighbors). Its appearance has less to do with details of the reionization process, but rather is a generic feature of any percolation process \citep{furlanetto16}. 

\subsection{Watershed Algorithm}  

The watershed algorithm is a well-known 2D image segmentation algorithm (see \citet{soille13}, and references therein). The essential idea is to connect constant values of a scalar field (e.g. greyscale), and treat them as contour lines indicating height in a topographic map. If one floods this topographic surface with water, it will break up into different catchment basins, demarcated by watershed lines. This provides the required image segmentation. We have extended the watershed algorithm to 3D. Such a 3D extension has been used for void-finding in both cosmological simulations and galaxy surveys \citep{platen07,neyrinck08,sutter15}. The detailed procedure is as follows: 
\begin{itemize} 
\item{{\bf Binary transform.} We first apply a threshold $x_{\rm HI}^{\rm th}$ such that  if the neutral fraction $x_{\rm HI} > x_{\rm HI}^{\rm th}$, we set $x_{\rm HI}=1$; otherwise, $x_{\rm HI}=0$. This eliminates ambiguity about whether a voxel is neutral or ionized. We choose $x^{\rm th}_{\rm HI} = 1$ (i.e., a voxel is ionized if and only if it is completely ionized); in practice we find we are not sensitive to this threshold.}

\item{{\bf Distance Transform.} Next, we assign to each ionized voxel $i$ the Euclidean distance $d_{i}$ to the nearest neutral voxel (if the voxel is neutral, then $d_{i}=0$). We use voxel units in which the side of a single voxel is one.} As previously noted, this 'distance transform' is identical to spherical averaging methods used by previous authors (\S\ref{section:spa}) in the limit of a two-phase medium. To ensure congruence with standard watershed terminology, we then invert $d_{i} \rightarrow - d_{i}$.

\item{{\bf Identify local minima.} We identify local minima of the scalar 3D array $d_{i}$. These naturally correspond to our notion of `bubble centers', since they are as distant as possible from HII region boundaries as possible. In watershed terminology, these are the bottoms of separate catchment basins. In principle, all that is now required is to identify the boundaries between different catchment areas (`watershed lines'). In practice, this leads to the well-known problem of over-segmentation, because every single local minimum, no matter how small, forms its own catchment basin. Thus, Poisson noise can lead to an obviously contiguous single bubble being spuriously sub-divided into many small bubbles.} 

\item{{\bf Threshold} To avoid this, we suppress shallow minima via the so called `h-minimum transform' that smooths out the map. Unlike Gaussian smoothing, such transformation preserves the topology of ionized regions without mixing neutral region voxels in it. Suppose a set of voxels $d_i$ rests in a catchment basin. In this catchment basin, there could be multiple sets of connected components (known as `markers' in image processing nomenclature) which create artificial sub-boundaries within a single catchment basin. The h-minimum transform smooths out those spurious local minima by imposing a contrast limit, $h$. If the differences in voxel value between a local minimum $d_{\rm min}$ and its surrounding voxels is less than $h$, then this shallow local minimum is potentially due to noise. We `cut off the tip' of this minimum by setting this voxel and all of its neighbors to the value $d_{\rm min}+h$ (note that $d_{\rm min}$ is negative, so this corresponds to reducing the depth of the minimum).
With an appropriate choice of $h$, we can eliminate small fluctuations at the bottom of the catchment basin which result in spurious over segmentation. Note that $h$ values we quote are in voxel units, not in physical units. Hence, the appropriate values of $h$ can change as we change the resolution of box.  }


\item{{\bf Identify bubble boundaries.} Now that the bubble centers have been found, we identify bubble boundaries. To do this, we inject fluid from markers (the minima of catchment basins) until they just touch. In practice, this corresponds to identifying contours of constant $d_{i}$. These `watershed lines' denote the bubble boundaries. In the absence of smoothing or thresholding, this procedure is guaranteed to assign every ionized voxel to a bubble. However, with thresholding, bubbles below the scale $h$ are either merged with larger bubbles or ignored. To conserve the total ionized fraction, we manually add back the smaller bubbles which were not merged into larger structures. These constitute a small fraction of the ionized volume ($\sim1\%$ of total ionized voxels ). In this paper, we use $h<1$. 
Thus, the thresholding process only affects minima in large bubbles (where neighboring voxels can have distances from the boundary which differ by less than unity). The smallest possible isolated bubble has $d_{\rm min}=1$, and is not eliminated by the smoothing process. Thus, this complication is not important for us.}

\end{itemize}

We now have a unique partitioning of ionized voxels into distinct bubbles, which we can use to calculate bubble volumes $V$. To compare against excursion set theory, we calculate the effective spherical radius $R_{\rm eff} = (3V/4 \pi)^{1/3}$ of each bubble. For a single spherical bubble, the watershed algorithm correctly identifies the entire region as ionized -- i.e., similar to FoF, its window function is a delta function at the true value. It also performs very well on Monte-Carlo catalogs, as we describe below.

\section{Results}
\label{section:results} 

\subsection{Monte-Carlo Tests}
\label{section:monte_carlo}

\begin{figure}
\begin{center}
\includegraphics[width=0.5\textwidth]{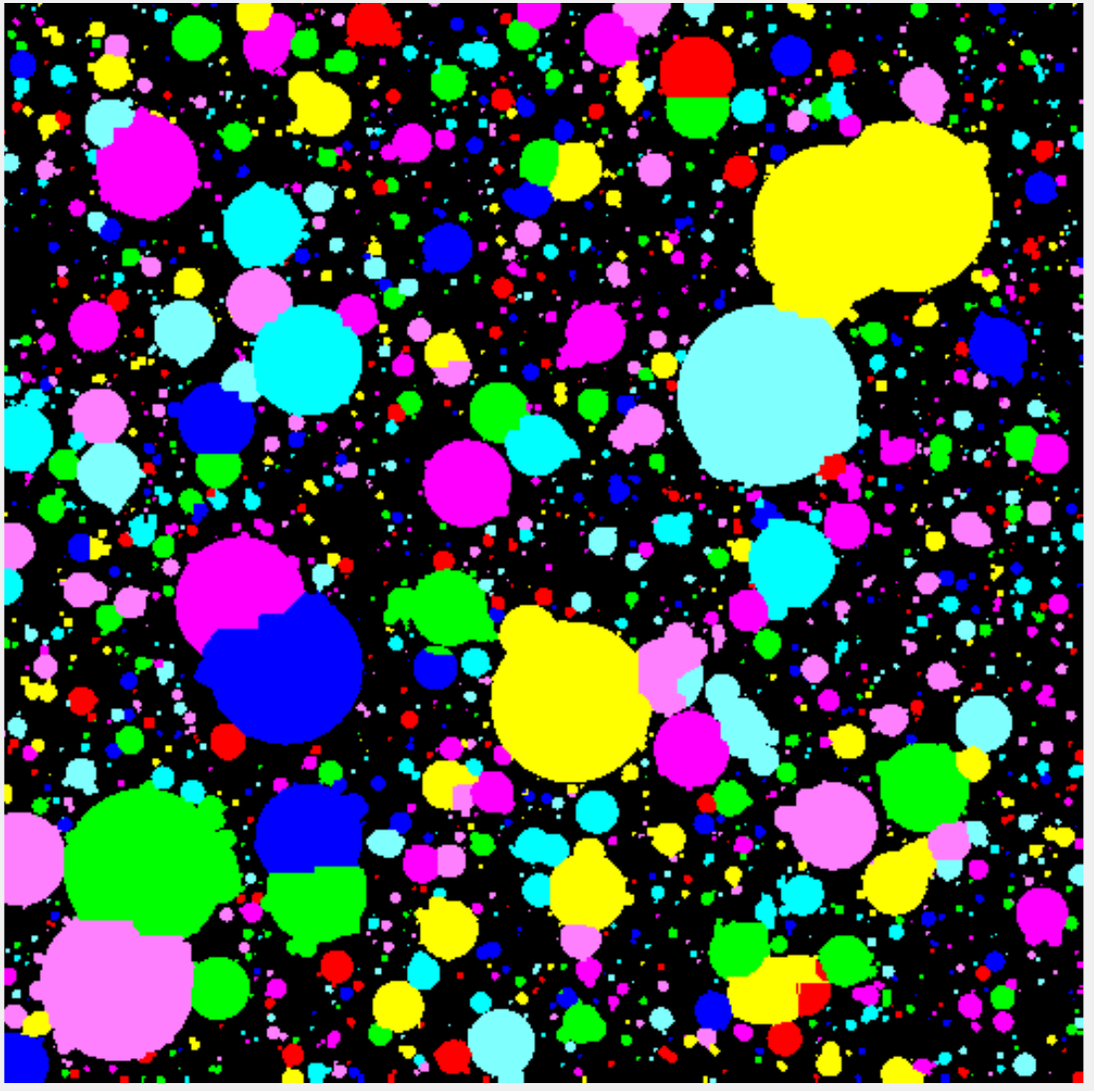}
\caption{A slice of the watershed segmentation of the Monte Carlo box; different colors denote separate bubbles.}
\label{fig:slice_monte}
\end{center}
\end{figure}

As a test bed and to better understand and validate our bubble-size measurement algorithms, we begin with Monte-Carlo bubble catalogs where the bubble size distribution is fully specified. We use the bubble size distribution given by excursion set theory; our fiducial test case corresponds to an ionization fraction of $Q_{\rm HII} = 0.50$ (and $z=11$, using an ionizing efficiency of $\zeta=40$ in the FZH04 model). We create a Monte-Carlo realization of spherical bubbles drawn from the FZH04 mass function in a $500^{3}$ box, 100 cMpc on a side. These bubbles are simply laid down randomly in a box; provided bubble overlap is small, we are guaranteed to be sampling from a known bubble PDF. This is {\it not} (as in semi-numeric algorithms, e.g., \citealt{mesinger07}) a direct implementation of the excursion set approach. As we shall see, the latter in fact has a bubble PDF quite different from that predicted by analytic theory.  

The results are shown in Fig \ref{fig:mfp_monte}. We show the probability density function $dP/d({\rm log \, R})$, normalized to integrate to unity. The results for each of the four methods are in line with what one might expect from the spherical window functions shown in Fig \ref{fig:window_functions}: (i) the watershed algorithm shows excellent performance, which is to be expected since it does an excellent job with isolated spherical bubbles (Fig. \ref{fig:window_functions}). The slightly narrower width can be understood from the adopted smoothing parameter ($h=0.7$; $h$ is in voxel units) which suppresses small bubbles. We explore sensitivity to smoothing in \S\ref{section:convergence}. (ii) The mean free path algorithm also does an excellent job. The slight shift toward smaller bubble sizes can be understood from the long tail of the window function toward small scales. Indeed the convolution of the excursion set PDF with the MFP window function closely corresponds to our results. (iii) By contrast, the distance transform performs very poorly in this Monte-Carlo trial, underestimating characteristic scales by an order of magnitude. Note that although the distance transform underestimates the single bubble size by a factor of $\sim 3$, once this window function is convolved with a bubble population, the distortion to the peak of the probability distribution can be considerably larger. The underlying excursion set PDF convolved with the DT window function matches our results. As one might expect, the watershed PDF convolved with the MFP and DT window functions also give excellent results. (iv) Finally, the FoF algorithm identifies most of the ionized volume in a single giant ionized bubble. The origins and implications of this result, which can be understood from percolation theory, are examined in detail in \citet{furlanetto16}.  

In Fig \ref{fig:slice_monte}, we show a slice through the simulation box. Each color labels a unique region found by the watershed algorithm. For the spatially Poisson distributed realization we have created, bubble overlap appears fairly minimal in 2D slices (note, however, that they are actually interconnected through narrow channels largely outside this slice, as can be shown using the FoF algorithm). In reality, bias at high redshift means that bubbles will inevitably be clustered \citep{furl04-bub}. Once bubble overlap is significant, we no longer know the underlying mass function -- indeed, its definition can be ambiguous (see \S \ref{section:discussion} for more discussion). As we shall see, clustering renders the watershed algorithm sensitive to the degree of smoothing applied.


\subsection{Application to Simulations of Reionization}
\label{section:real_sims}

\begin{figure}
\begin{center}
\includegraphics[scale = 0.6]{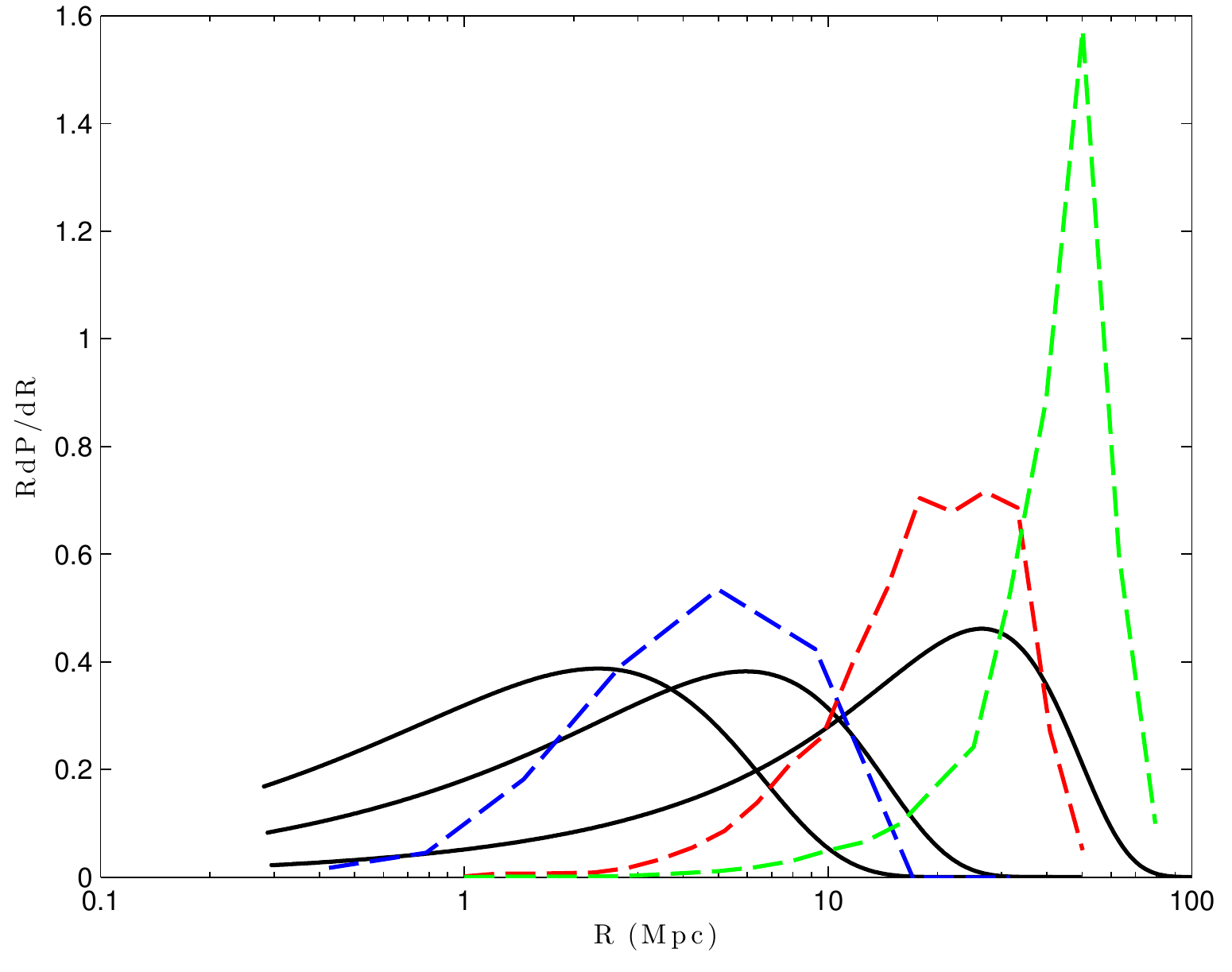}
    \caption{Watershed bubble PDFs of semi-numeric simulation boxes with $Q_{\text{HII}} = 0.28,0.49,0.78$ ($\zeta = 55, 54, 50$ at $z=12,11,10$). The black solid lines from left to right are the excursion predictions for increasing $Q_{\text{HII}}$, while the dashed lines from left to right are the watershed results.}
    \label{fig:water_all}
\end{center}
\end{figure}

\begin{figure}
\begin{center}
\includegraphics[scale = 0.6]{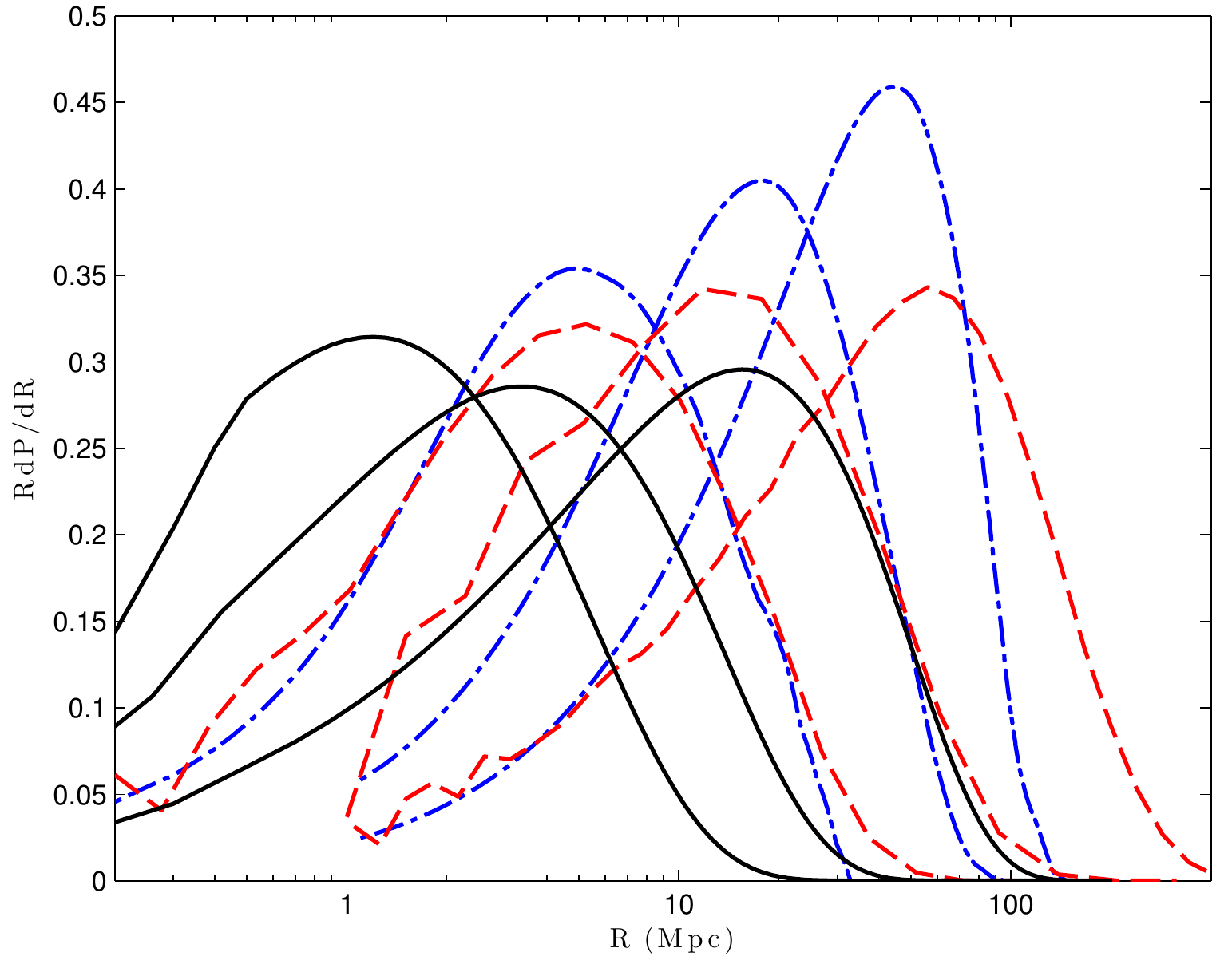}
    \caption{Mean free path (MFP) PDFs of semi-numeric simulation boxes with $Q_{\text{HII}} = 0.28,0.49,0.78$ ($\zeta = 55, 54, 50$ at $z=12,11,10$). The black solid lines from left to right are the excursion predictions convolved with the MFP window function, in order of increasing $Q_{\text{HII}}$. The blue dot dashed lines are the watershed results (also convolved with the MFP window function), and the red  dashed lines are the mean free path results with $10^7$ samples.}
    \label{fig:mfp_all}
\end{center}
\end{figure}


\begin{figure}
\begin{center}
\includegraphics[scale = 0.6]{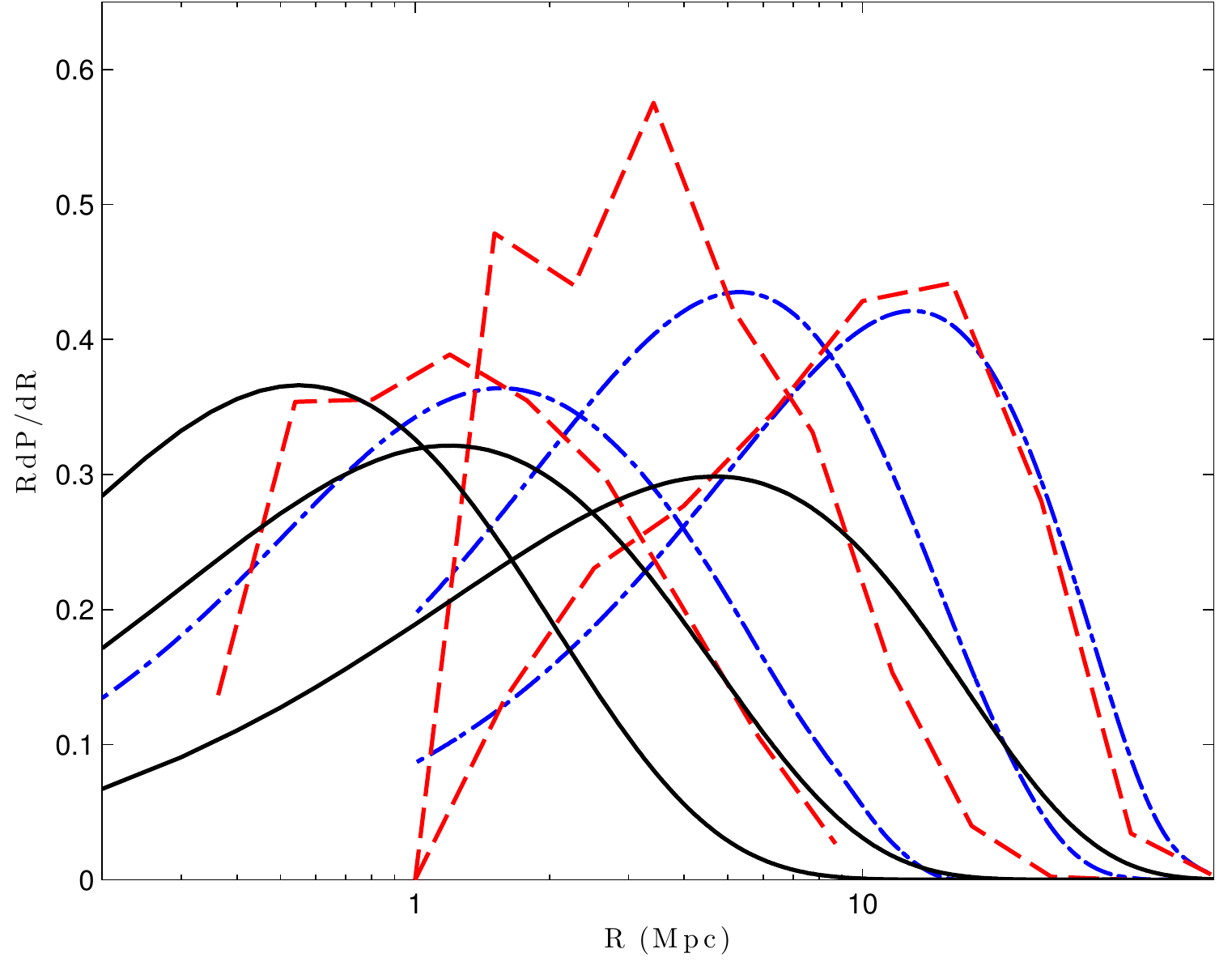}
    \caption{Distance transform probability density functions for semi-numeric simulation boxes with $Q_{\text{HII}} = 0.28,0.49,0.78$ ($\zeta = 55, 54, 50$ at $z=12,11,10$). The black solid lines from left to right are the excursion set predictions convolved with the DT window function, in order of increasing $Q_{\text{HII}}$. The blue dot dashed lines are the watershed results (also convolved with the DT window function), and the red  dashed lines are results for the direct distance transform.}
    \label{fig:dist_all}
\end{center}
\end{figure}

\begin{figure}
\begin{center}
\includegraphics[width=0.5\textwidth]{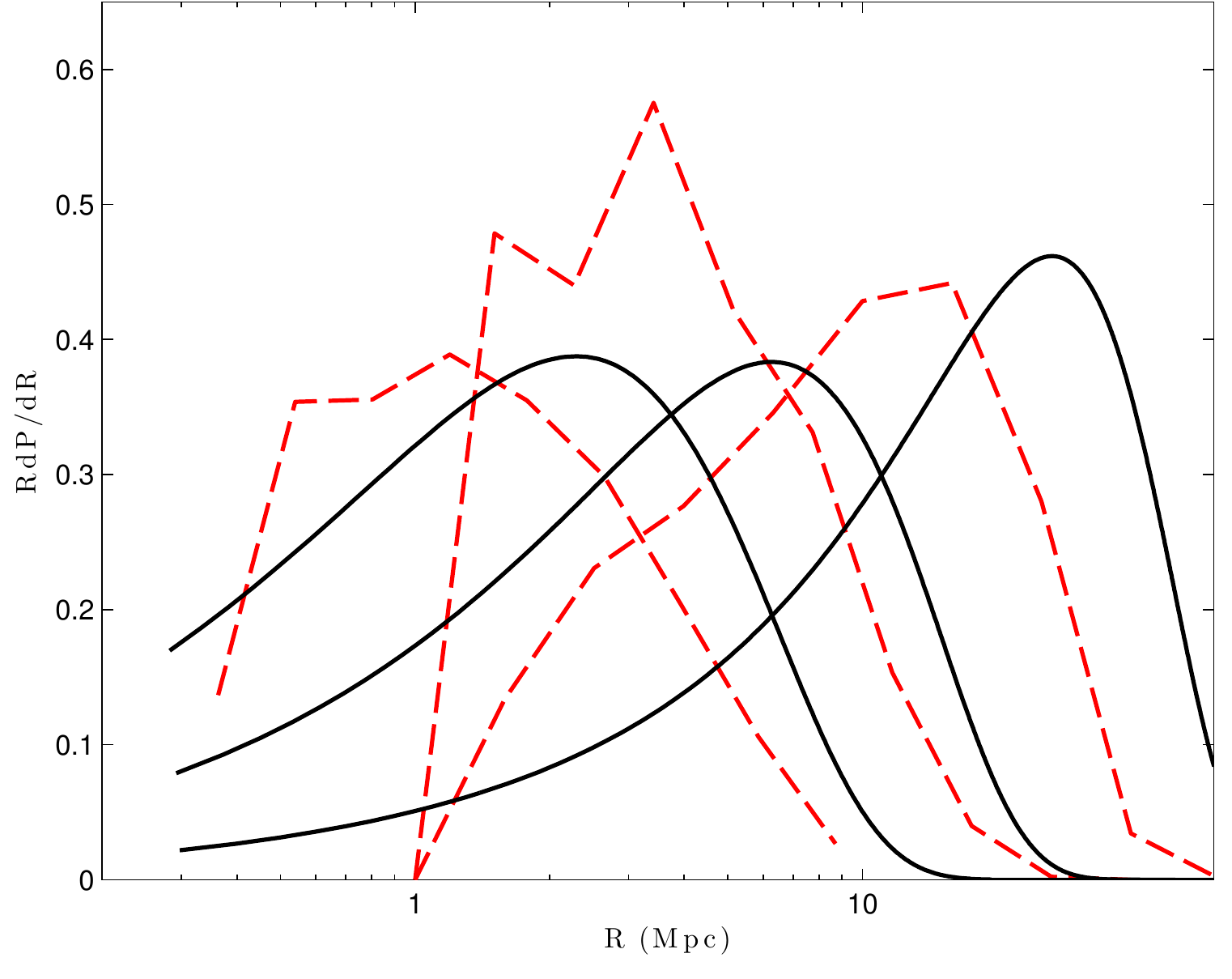}
    \caption{Distance transform probability density functions for semi-numeric simulation boxes versus the excursion set predictions ({\it not} convolved with the DT window function). The black lines are excursion set predictions and the red curves are distance transform results at $z=12$, $11$, and $10$ ($Q_{\text{HII}} = 0.28,0.49,0.78$) from left to right respectively. These show apparent (and spurious!) agreement. The cutoff at 1 Mpc (and consequent distortion of the PDF) for the $z=11,10$ boxes is due to finite resolution effects; see discussion in text.}
    \label{fig:dist_vs_unconv}
\end{center}
\end{figure}

Now that our bubble size measurement algorithms have been carefully characterized and tested, we deploy them on semi-numeric simulations of reionization. We use the 21cmFAST code \citep{mesinger07},  which allows large, relatively high resolution simulations to be computationally affordable, by obviating the need for hydrodynamics or radiative transfer. We shall find that accurate resolving the bubble PDF, particularly in the late stages of reionization, requires large boxes ($\sim 500$ Mpc), which are difficult with numerical simulations due to the large required dynamic range. While they may differ in detail from radiation-hydrodynamics simulations, the gross morphology of HII regions in side-by-side comparisons of the two approaches starting from the same initial conditions is extremely similar \citep{mcquinn07,zahn07,zahn11}. For our purposes, where we aim to explore the gross evolution of the characteristic scale with redshift, rather than fine details of the bubble PDF, this level of accuracy is sufficient. The main limitation with standard semi-numeric simulations is that they incorporate opacity due to Lyman limit systems in a very approximate manner. The latter can change the morphology of the ionization field, particularly in the late stages of reionization \citep{furl05-rec,finlator12,sobacchi14}. Thus, reported bubble sizes for the late stages of reionization may be less reliable. In any case, as $Q_{\rm HII} \rightarrow 1$, it is probably more sensible to construct a size PDF of {\it neutral} regions. Also, it should be noted that the `mean free path' here refers to bubble sizes, and does not take into account the fact that photon mean free paths are eventually regulated not by bubble sizes but by Lyman limit systems. 

The 21cmFAST code uses an excursion-set theory approach to generating the ionization field\footnote{It may seem wrong-headed to use excursion-set theory simulations to test excursion set theory. We discuss this point in \S\ref{section:discussion}.}. A linear density field is generated at high redshift, then evolved to lower redshift using linear theory and the Zel'dovich approximation. Using a conditional mass function from a voxel's mean density, a source catalog is generated for halos with virial temperature $T> 10^{4}$K (as needed for hydrogen atomic cooling). The ionization field is then generated from the source catalog using excursion set theory (specifically, finding the largest spherical regions which can be ionized by the enclosed sources). We use the default setting of a maximum bubble size of $R_{\rm max}=30$ Mpc to crudely model the effects of absorption of ionizing photons in the IGM\footnote{This setting in 21CMFAST sets the maximum size of a single spherical ionized region which can be obtained using the spherical filtering procedure. This does not directly correspond to the bubbles given by our procedure --- for instance, if two 30 Mpc bubbles overlap, this gives rise to a larger ionized region.}. Our boxes are $500^{3}$. We use a cosmology of $\Omega_{\rm m}=0.31$, $\Omega_{\Lambda}=0.69, \ \Omega_{b}=0.049, \ h=0.67, \ n=0.96, \sigma_{8}=0.83$ that matches the results of Planck 2015 \citep{planck-collaboration15}. We examine bubble PDFs at three $Q_{\rm HII}$ in simulations boxes with the properties given in Table \ref{tab:models}. We use the ionization efficiency $\zeta = 55, 54, 50$ at $z=12,11,10$ to match the values of $Q_{\rm HII}$ to the true ionization fraction as found in our simulation boxes. Note that we increase the physical size of the simulation boxes at lower redshifts, when $Q_{\rm HII}$ and bubble sizes are larger. Failure to do so results in considerable cosmic variance, due to the small number of large bubbles which consume most of the volume. We discuss this point in more detail later. 


\begin{table}
\begin{center}
\caption{Simulation Parameters}
\begin{tabular}{ c|c|c|c } 
 \hline
 \hline
 $z$ & $Q_{\rm HII}$& Box Size &L (Mpc)\\
 \hline
 12 & 0.28 & $500^{3}$ &100 \\ 
 11 & 0.49 & $500^{3}$ &500\\ 
 10 &0.78 & $500^{3}$ &500\\ 
 \hline
\end{tabular}\label{tab:models}
\end{center}
\end{table}

In Fig \ref{fig:water_all}, \ref{fig:mfp_all}, \ref{fig:dist_all}, we compare the bubble PDF predicted by excursion set theory against the PDF recovered from the simulations by the watershed algorithm, the MFP method, and the distance transform respectively. Several features are immediately apparent. For all methods, the bubble size distribution is peaked toward larger bubble sizes than excursion set theory, by up to an order of magnitude. The difference is particularly egregious during the early stages of reionization. The simulations also show bubble sizes which evolve less rapidly during reionization than predicted by excursion set theory. Importantly, these findings only hold when theoretical predictions are convolved with the window function of the appropriate method. If one uses unconvolved theoretical predictions--thus failing to compare apples with apples-- apparent spurious agreement between excursion set theory and simulations can arise, a problem which bedeviled previous comparisons. We can see this in Fig \ref{fig:dist_vs_unconv}, which shows good apparent agreement between the DT and analytic excursion set theory, if one does not convolve with the appropriate window function.\footnote{The agreement is somewhat poorer for the DT at $Q_{\rm HII}=0.49$ ($z=11$). This is an artifact of the lower resolution we have adopted for this case ($\Delta x=1$ Mpc), which truncates the bubble PDF for small bubble sizes; agreement would almost certainly improve for higher resolution. This issue for our choice of box sizes only arises for the distance transform, which incorrectly underestimates bubble sizes and therefore appears to require higher resolution. Since the DT is at any rate not an accurate method, we do not pursue further refinements.} 

Figs \ref{fig:mfp_all} \& \ref{fig:dist_all} show that when the watershed PDFs are convolved with the appropriate window functions for the mfp and distance transform methods, they match the direct MFP and distance transform measurements well.\footnote{Note again the poorer agreement for the DT at $Q_{\rm HII}=0.49$, $z=11$, for the reasons previously mentioned.} This gives us further confidence that the watershed algorithm accurately measures the true underlying bubble PDF. It also suggests that the window function we have adopted for MFP and distance transform methods (which strictly is only for spherical bubbles) is still broadly applicable. 

The watershed PDFs are noticeably narrower than the excursion set PDFs at all stages of reionization; in particular, they lack an extended tail toward small bubble sizes, even after accounting for resolution effects. Closer examination of the simulations reveals that many of the small bubbles predicted by excursion set theory have merged together to create larger bubbles, early in the reionization process. The watershed PDFs also become increasingly narrow as reionization progresses. 

\begin{figure}
\begin{center}
\includegraphics[width=0.5\textwidth]{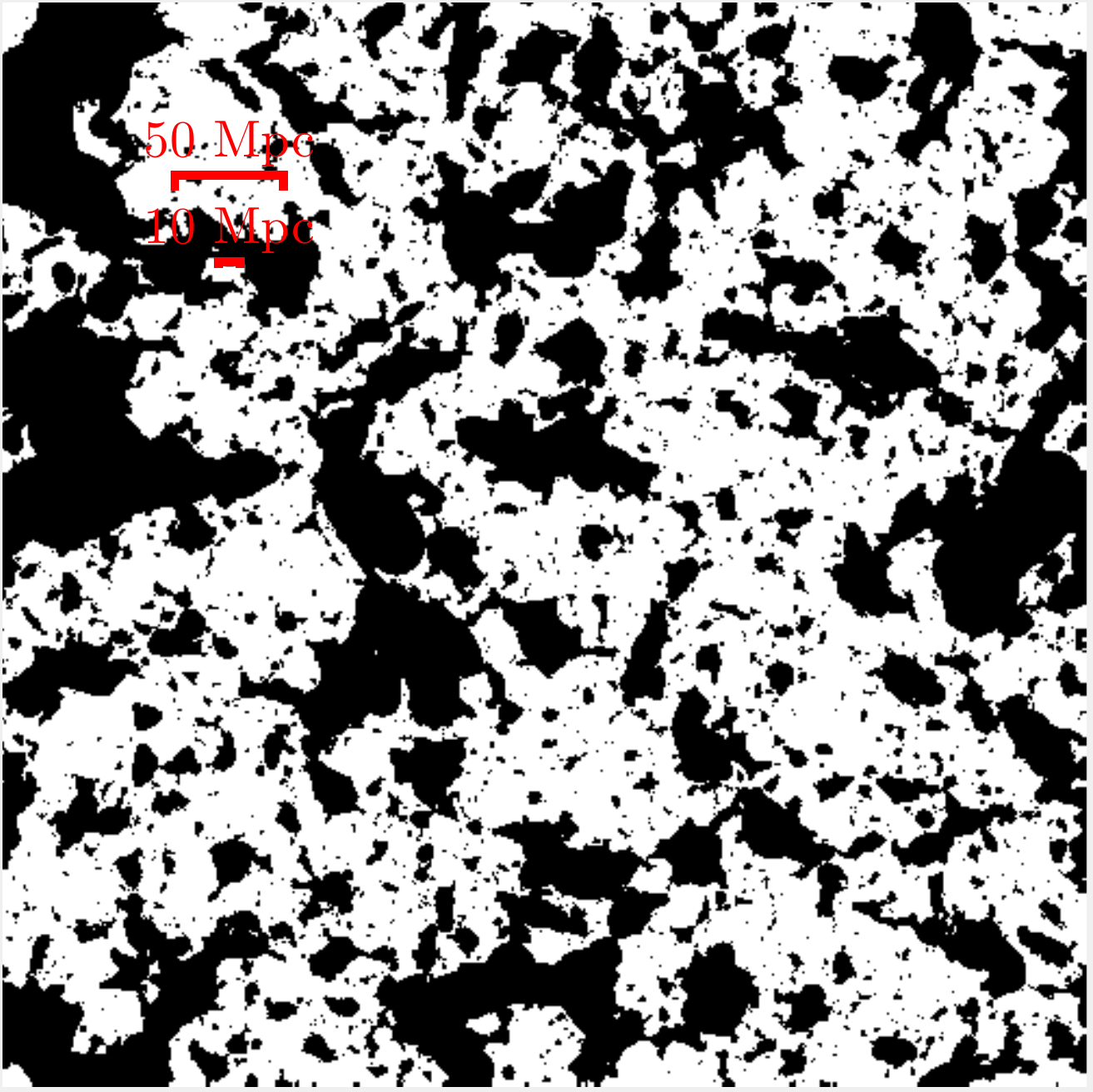}
\includegraphics[width=0.5\textwidth]{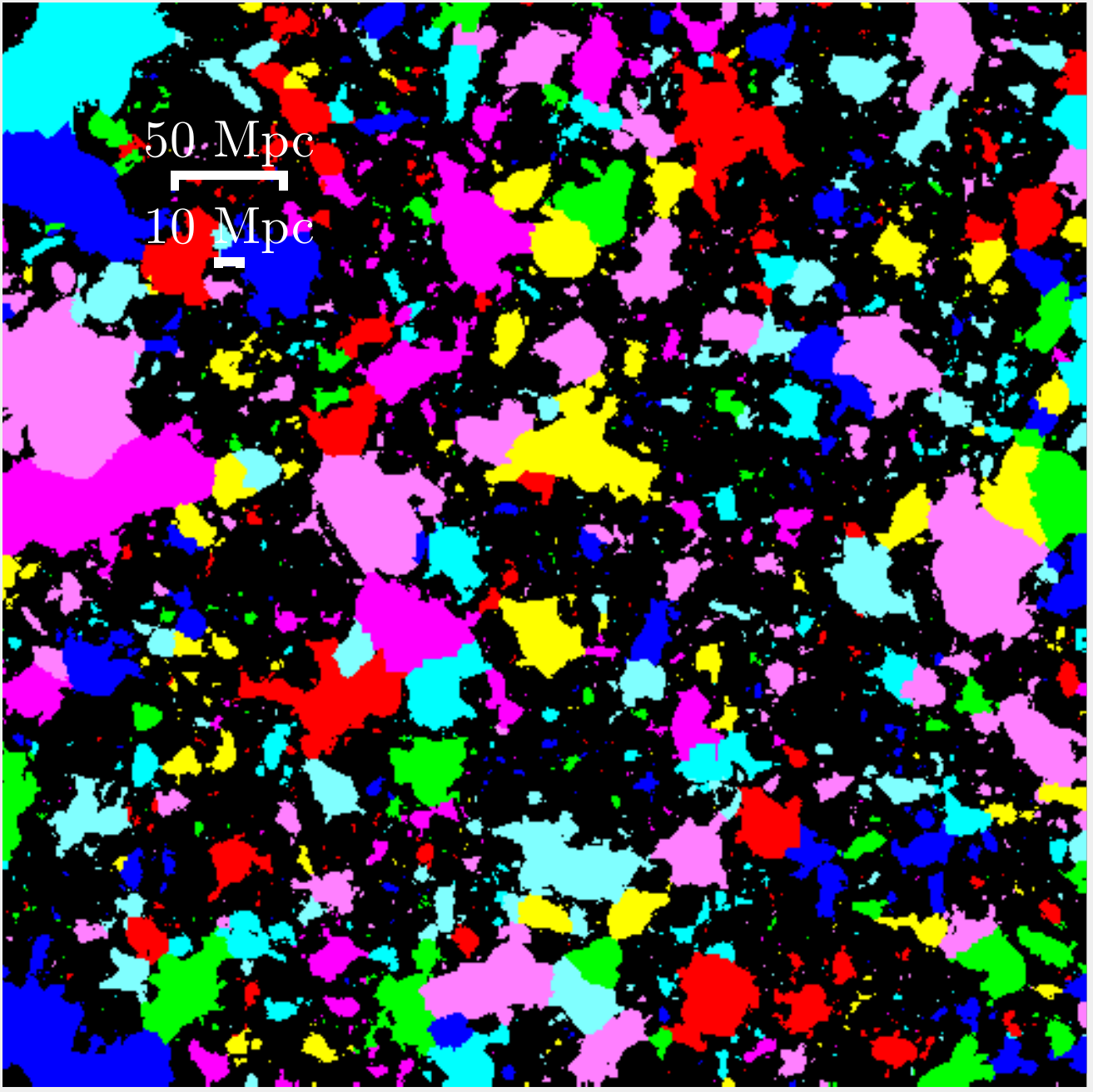}
    \caption{Top panel: slice of a simulation box with characteristic scales predicted by excursion set and uncovered by watershed methods at $Q_{\text{HII}} = 0.49$ ($z=11$). The excursion set has characteristic diameter of about 10 Mpc (10 voxels) and the watershed has characteristic diameter of about 50 Mpc (50 voxels) as shown. Bottom panel: the same, but showing the watershed segmentation.}
    \label{fig:scales}
\end{center}
\end{figure}

\begin{figure}
\begin{center}
\includegraphics[scale=0.3]{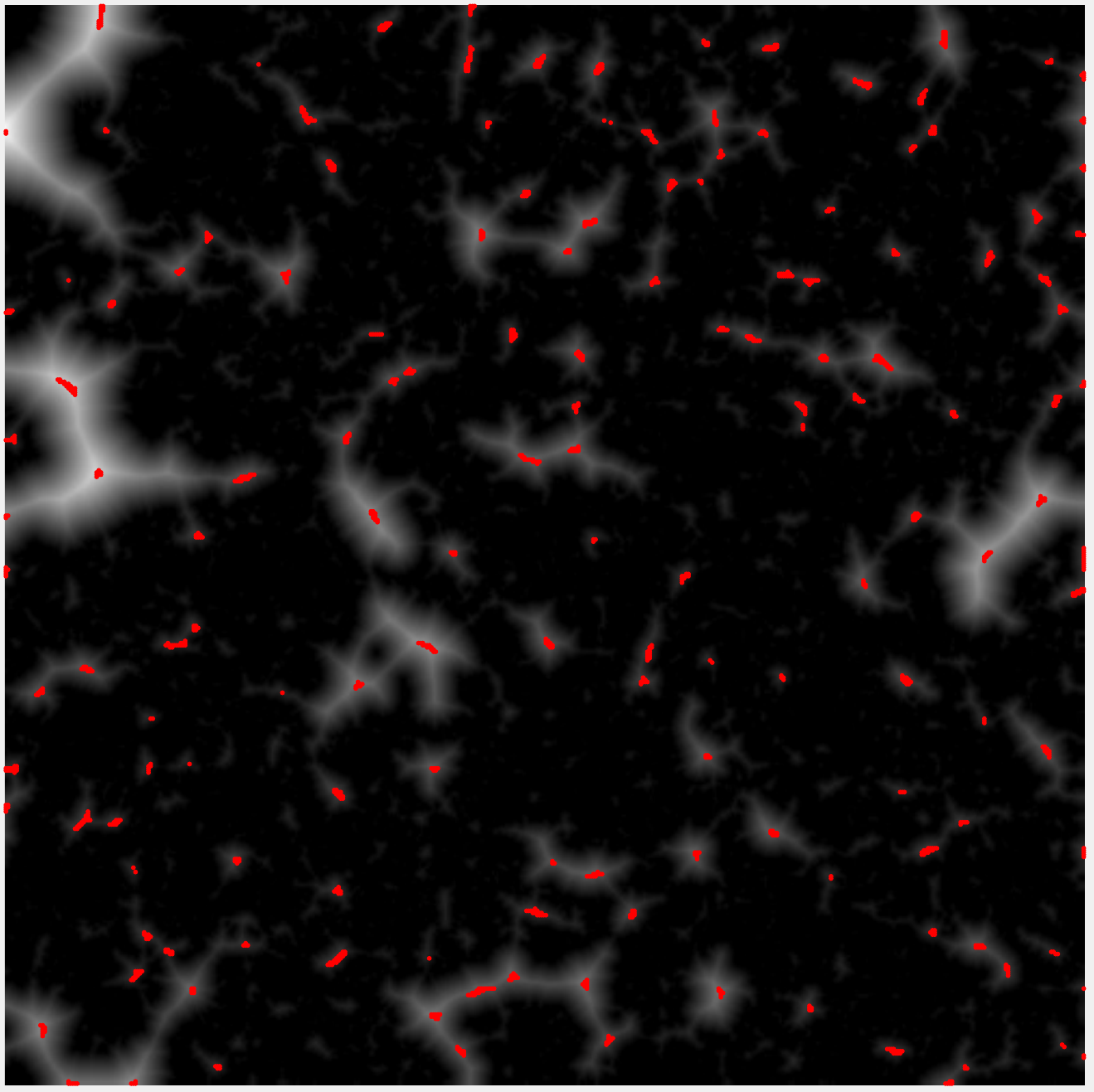}
\includegraphics[scale=0.36]{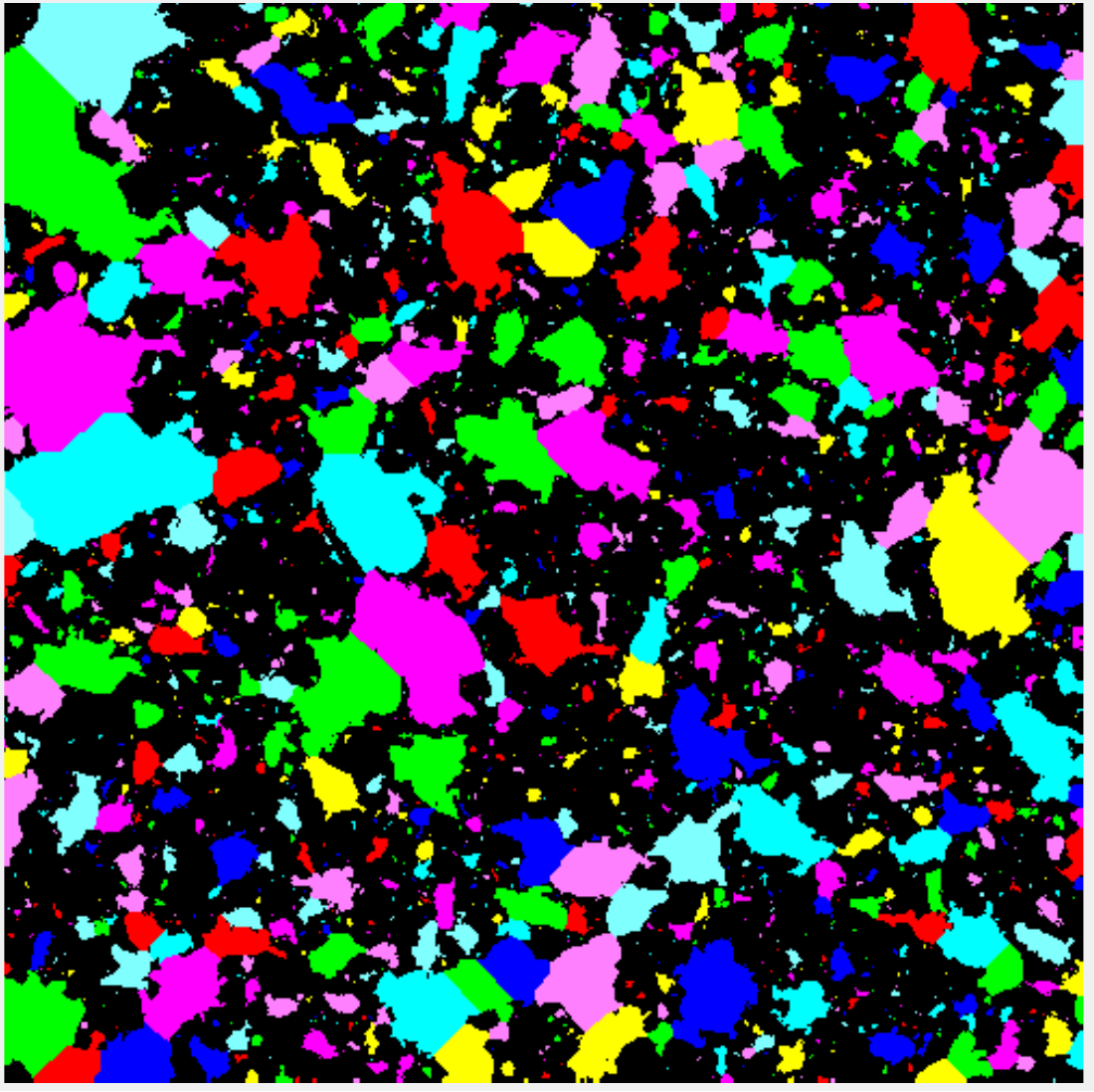}
\includegraphics[scale=0.3]{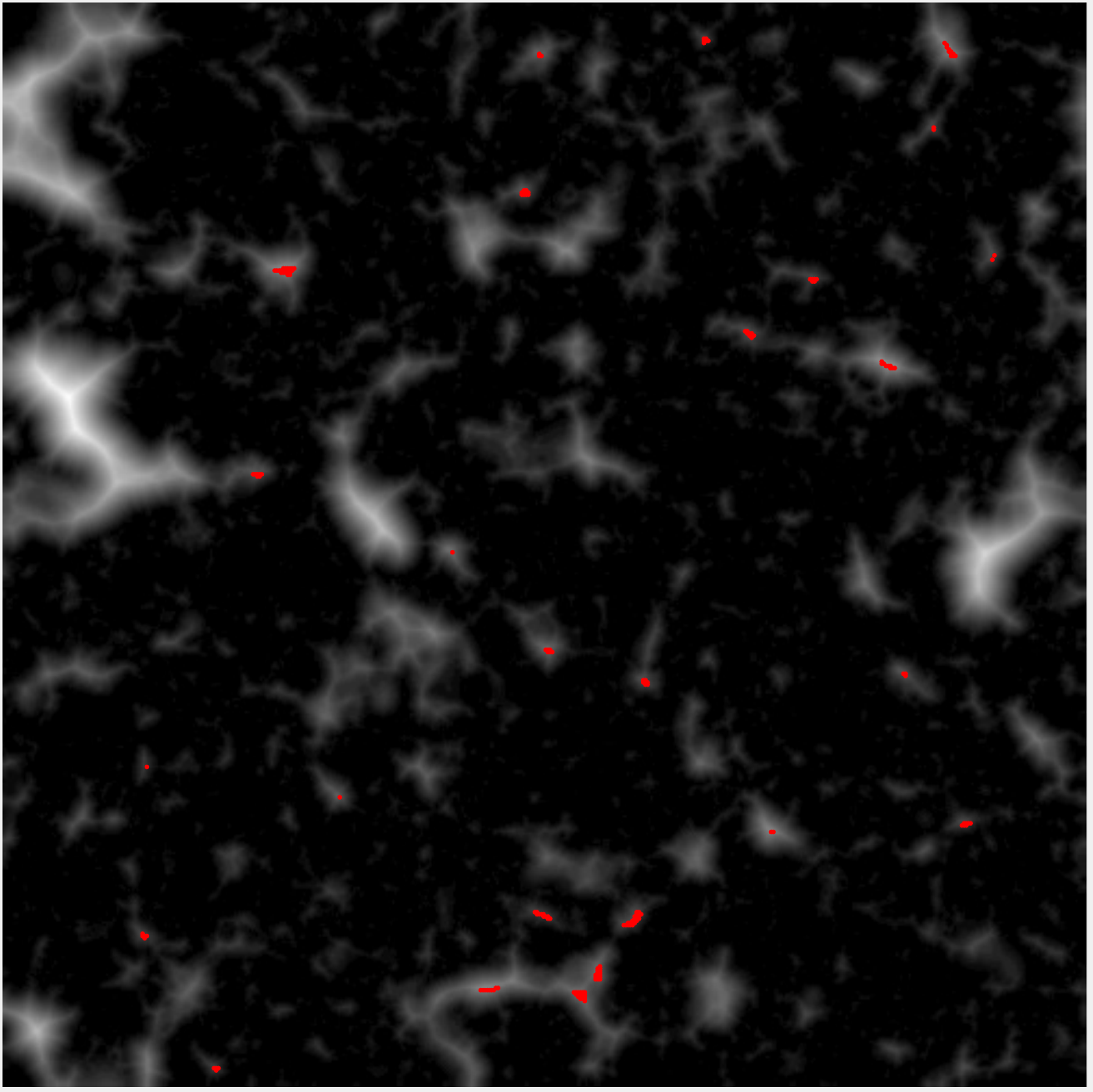}
\includegraphics[scale=0.36]{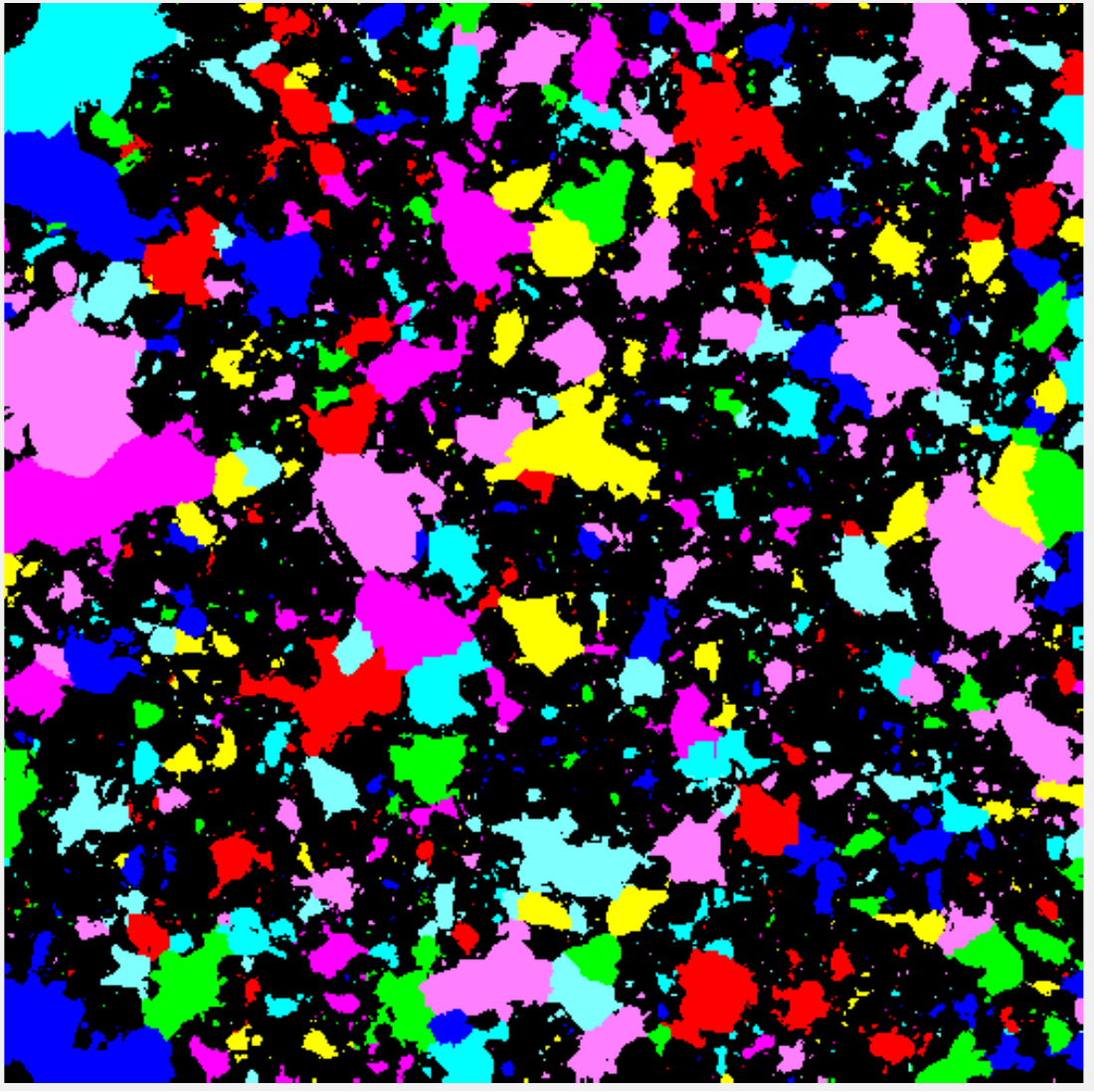}
    \caption{Top row: bubble centers (left) and segmentation (right) from the watershed algorithm applied to a chosen 2D slice at $z=11$ and smoothing parameter $h=0.9$. Bottom row: bubble centers (left) and segmentation (right) slices of the full 3D watershed algorithm at same redshift and smoothing parameter. Bubble centers and segmentation are significantly different for the 3D case. Note that we have shown the bubbles with distance transform value greater than 5 pixel/voxel units for each slice with red markers. In reality, there are many more smaller bubbles in both slices.}
    \label{fig:markers}
\end{center}
\end{figure}

In the upper panel of Fig \ref{fig:scales}, we show slices of simulation boxes with characteristic scales predicted by excursion set theory and the watershed algorithm shown. It is immediately visually apparent that the excursion set theory predictions are too small-- no sophisticated methods are required! By contrast, the scales uncovered by the watershed algorithm appear quite sensible. The segmentation produced by the watershed algorithm is also visually sensible (lower panel). There are some cases where the segmentation may seem somewhat arbitrary (e.g., a large contiguous bubble is segmented into smaller pieces). Upon further examination, this is almost invariably due to 3D structure not visible in 2D slices. We caution the reader that 2D slices-- from which much of our intuition has been drawn over the years-- can be misleading. For instance, 2D circles are unlikely to be portions of spheres, but instead part of more complicated features which can only be fully understood in a 3D rendering. We emphasize that all algorithms need to be fully 3D, and furthermore 3D visualizations are very important for building intuition. To drive this point home, in Fig \ref{fig:markers}, we show the results of the watershed algorithm as applied to a 2D slice, and then the results for that same slice when it is incorporated into a full 3D algorithm. The results of the segmentation process are different. One of the reasons is clear from the left panel. The red points indicate bubble centers (i.e., minima of the distance transform after h minimum smoothing). The reader should focus on the largest ionized regions (the plethora of red dots outside are due to tiny ionized regions). In a 2D segmentation, there are always a few red points per large ionized region, whereas in the 3D algorithm, often there are none: the bubble centers lie elsewhere, once the 3D topology is taken into account. 

In Fig 11, we show 3D renderings of the 5 largest bubbles in a $L=100$ Mpc portion (chosen for the sake of clarity) of the $Q_{\rm HII}=0.49$ ($z=11$) box. These largest bubbles are reasonably symmetric and not too complex in their topology, as one might expect for volume-filling regions. This is to be contrasted with the complex network of tunnels uncovered by a friends-of-friends or percolation algorithm (see corresponding figures in \citealt{furlanetto16}). 



\begin{figure}
\begin{center}
\includegraphics[width=0.5\textwidth]{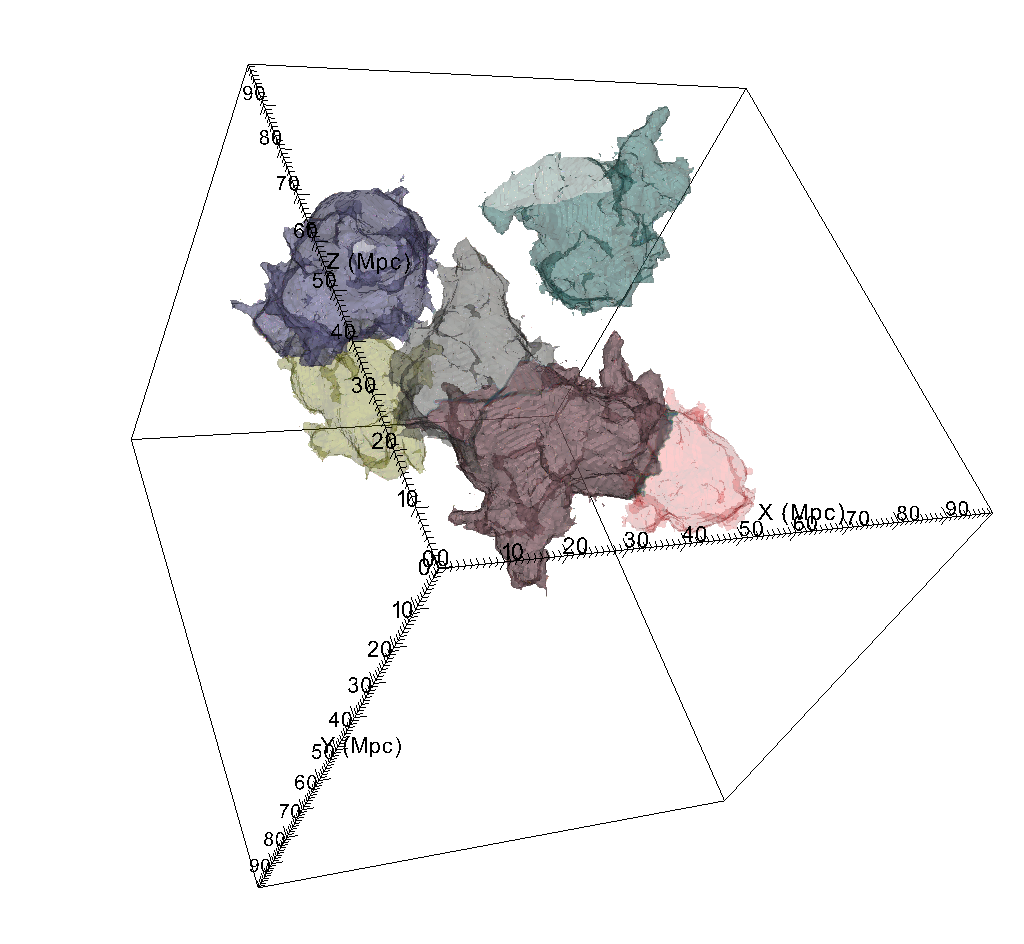}
    \caption{5 biggest bubbles found by the watershed algorithm, in a 3D rendering of a 100 Mpc region of the $Q_{\rm HII}=0.49$ $(z=11)$ box.}
    \label{fig:3D}
\end{center}
\end{figure}

\section{Convergence tests}
\label{section:convergence}  

There are two important areas where we need to check convergence for the watershed algorithm, which has not been applied before in this context. Firstly, the effects of box size and resolution. Secondly, the watershed algorithm has an adjustable smoothing parameter $h$ ($h$ is quoted in voxel units--e.g., $h=2$ corresponds to 2 voxels). It is important to check sensitivity to this parameter. As with any regularization scheme, the smoothing parameter has a critical effect on the output, and the optimal choice is not always clear. 

\begin{figure}
\begin{center}
\includegraphics[scale = 0.6]{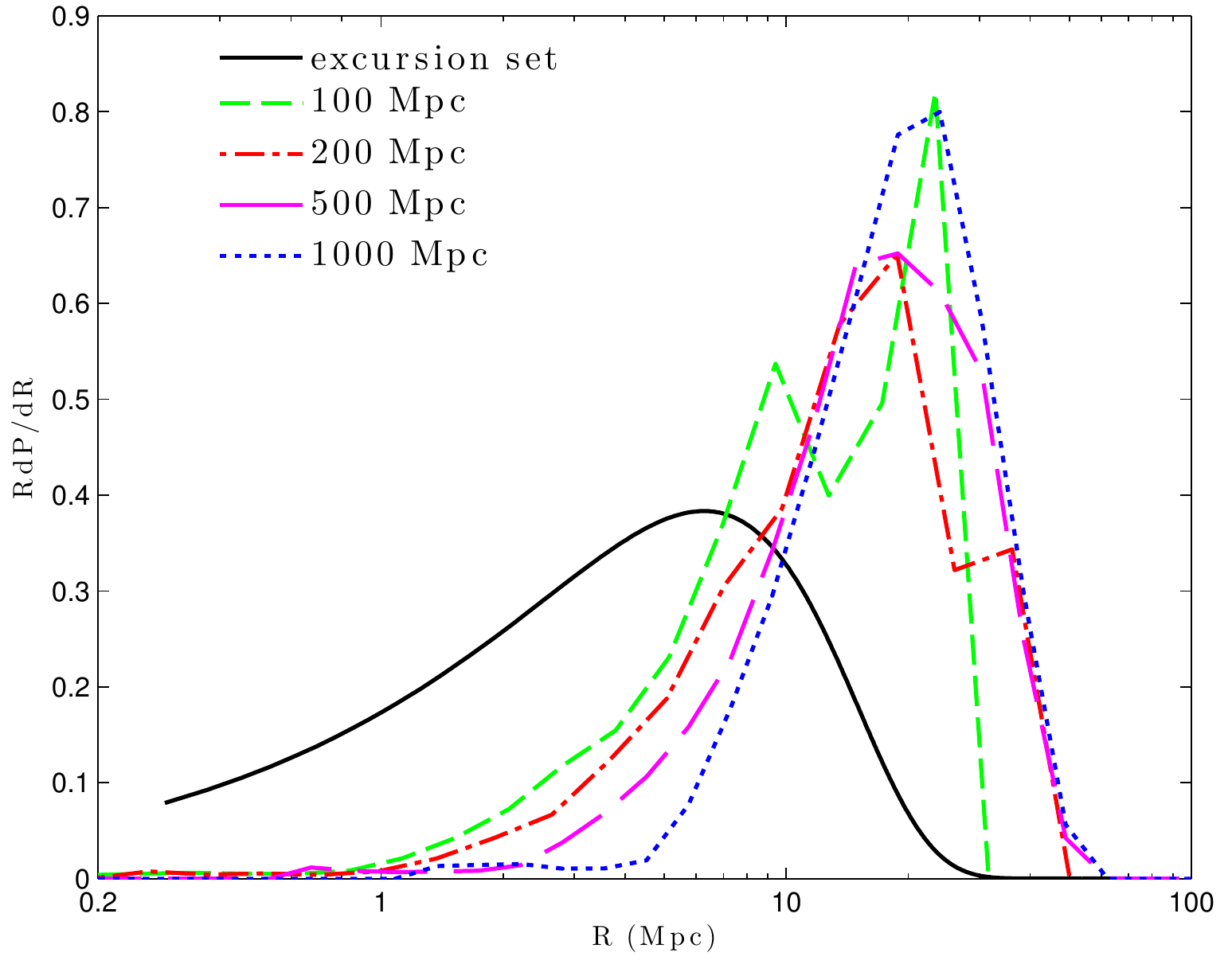}
    \caption{Watershed bubble PDFs of semi-numeric simulation boxes with smoothing parameter $h=0.7$ and $Q_{\text{HII}} \approx 0.48$ ($z\approx11$), for various box sizes $L$.  $Q_{\text{HII}}, z$ are approximate since the ionization fraction changes slightly as the scales are changed. We adjust $z$ accordingly to account for such differences. Each box is $500^{3}$. The black solid line is the excursion set prediction, the green short dashed line, the red dotted dashed line, the pink long dashed line and the blue dotted line correspond to $100,200,500,1000$ Mpc respectively.}
    \label{fig:convergent_z12}
\end{center}
\end{figure}

First, we examine the tradeoff between box size and resolution, for our fixed $500^{3}$ boxes. If $\Delta x, R_{b}, L$ are the voxel size, characteristic bubble size, and box size respectively, then ideally we should have $\Delta x \ll R_{b} \ll L$. In Fig \ref{fig:convergent_z12}, we examine the convergence properties of the watershed bubble PDF as the box size $L$ is changed (and hence $\Delta x=L/N$, where we have used $N=500$ throughout). The PDF--particularly the location of the PDF peak--is fairly stable to box size/resolution for $L/R_{\rm b}=5,10,25,50$ (and thus $R_{\rm b}/\Delta x=100,50,20,10$). However, two interesting features are present. Firstly, in the $L=100$ Mpc ($L/R_{\rm b}=5$) simulation, the bubble PDF is bimodal! We have seen this in simulations with different $Q_{\rm HII}$ (and $z$) and resolution for small box sizes. It is consistent with Poisson fluctuations in the number of rare large bubbles (which consume most of the volume) in a small box. For an appropriately normalized bubble PDF, as the number of rare large bubbles fluctates, so does the relative contribution of smaller bubbles. We conclude that as rule of thumb, $L/R_{b} \gsim 10$ is needed. On the other hand, bubble detection appears to be suppressed for bubbles with $R_{\rm min} \lsim 3 \Delta x$. The bubble PDFs in the semi-numeric simulations are sharply peaked (and become increasingly so as reionization progresses). Typically there is no significant fractional contribution from bubbles with $R_{\rm min} < R_{b}/10$. We conclude that $R_{b}/\Delta x \gsim 30$ is a reasonable rule of thumb. Thus, requiring $L/R_{b} \gsim 10$, $R_{b}/\Delta x \gsim 30$ implies that $300^{3}$ simulations are potentially adequate. Of these two requirements, it is more important to satisfy the box size threshold $L/R_{\rm b} \gsim 10$, since there is relatively little volume in the small bubbles. These considerations drive our choice of box sizes shown in Table \ref{tab:models}. 

\begin{figure}
\begin{center}

\includegraphics[width=0.3\textwidth]{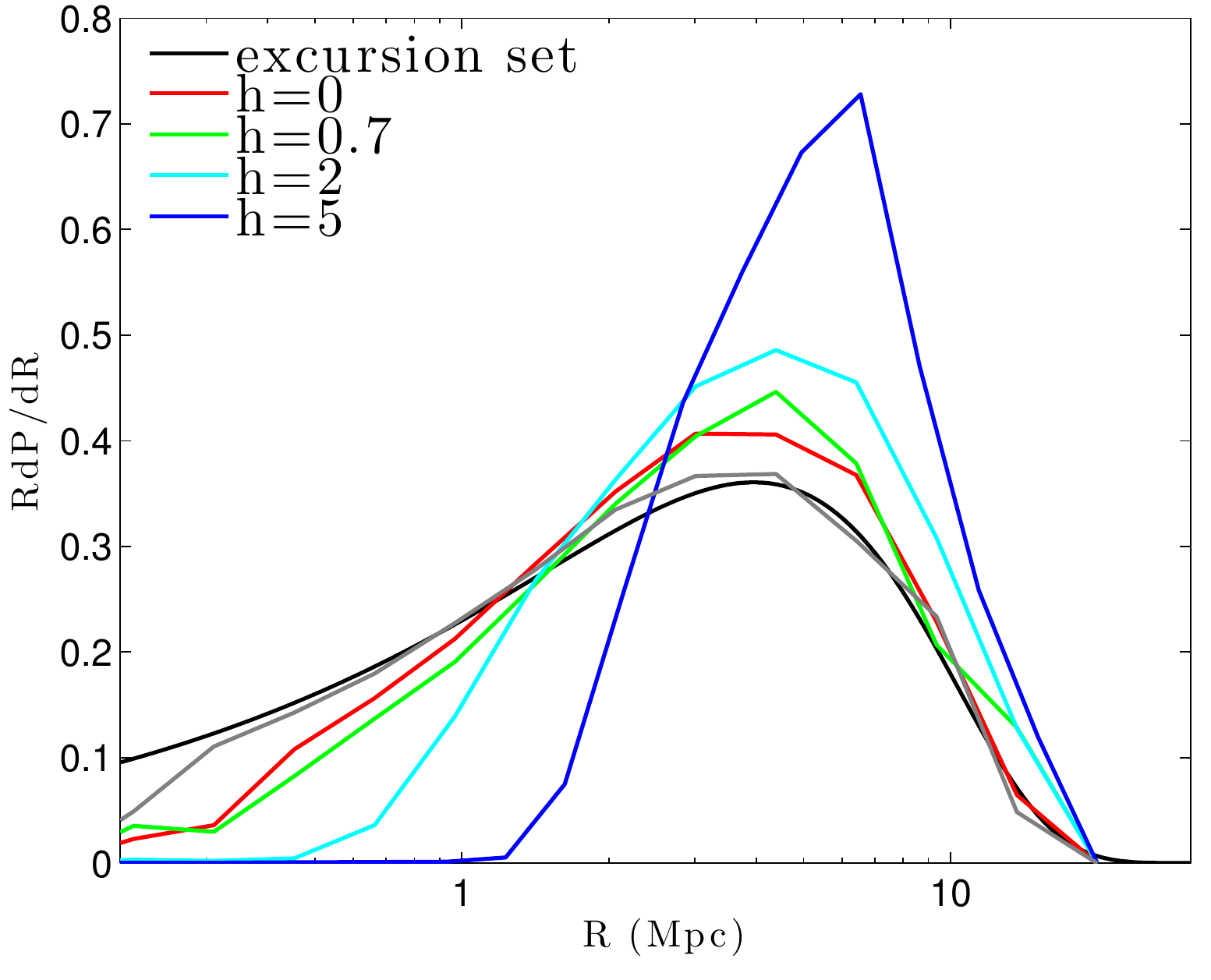}
\includegraphics[width=0.3\textwidth]{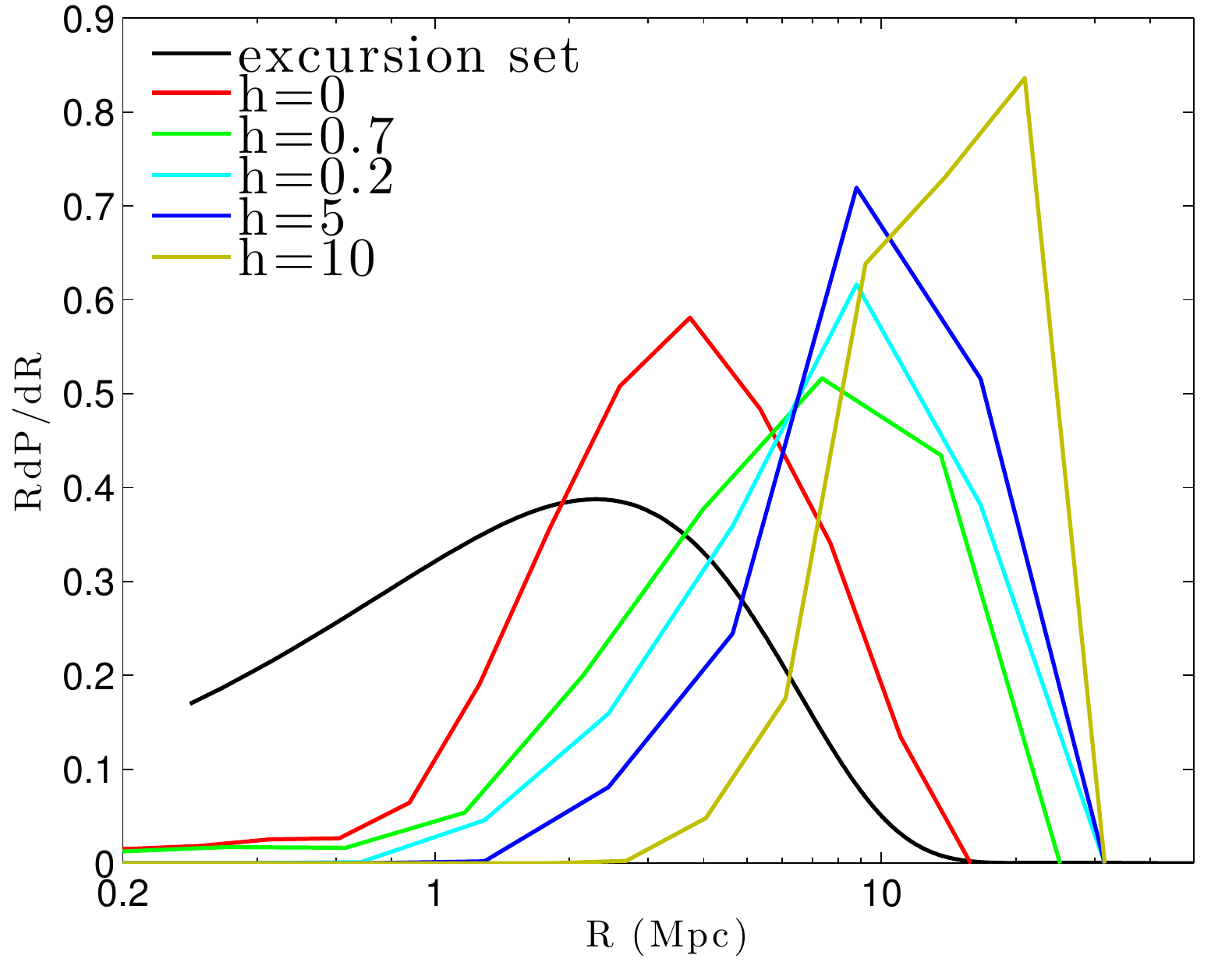}
\includegraphics[width=0.3\textwidth]{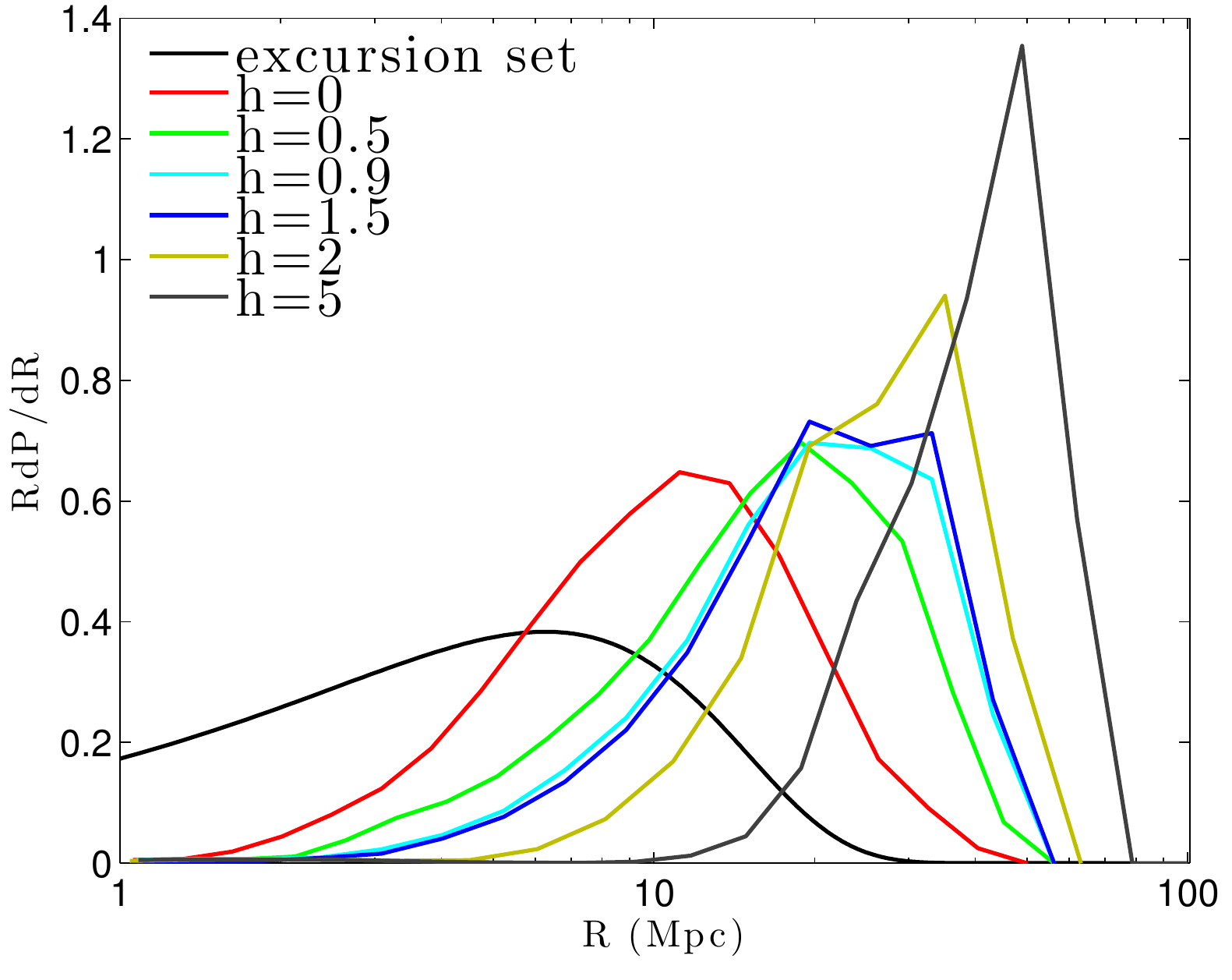}
\includegraphics[width=0.3\textwidth]{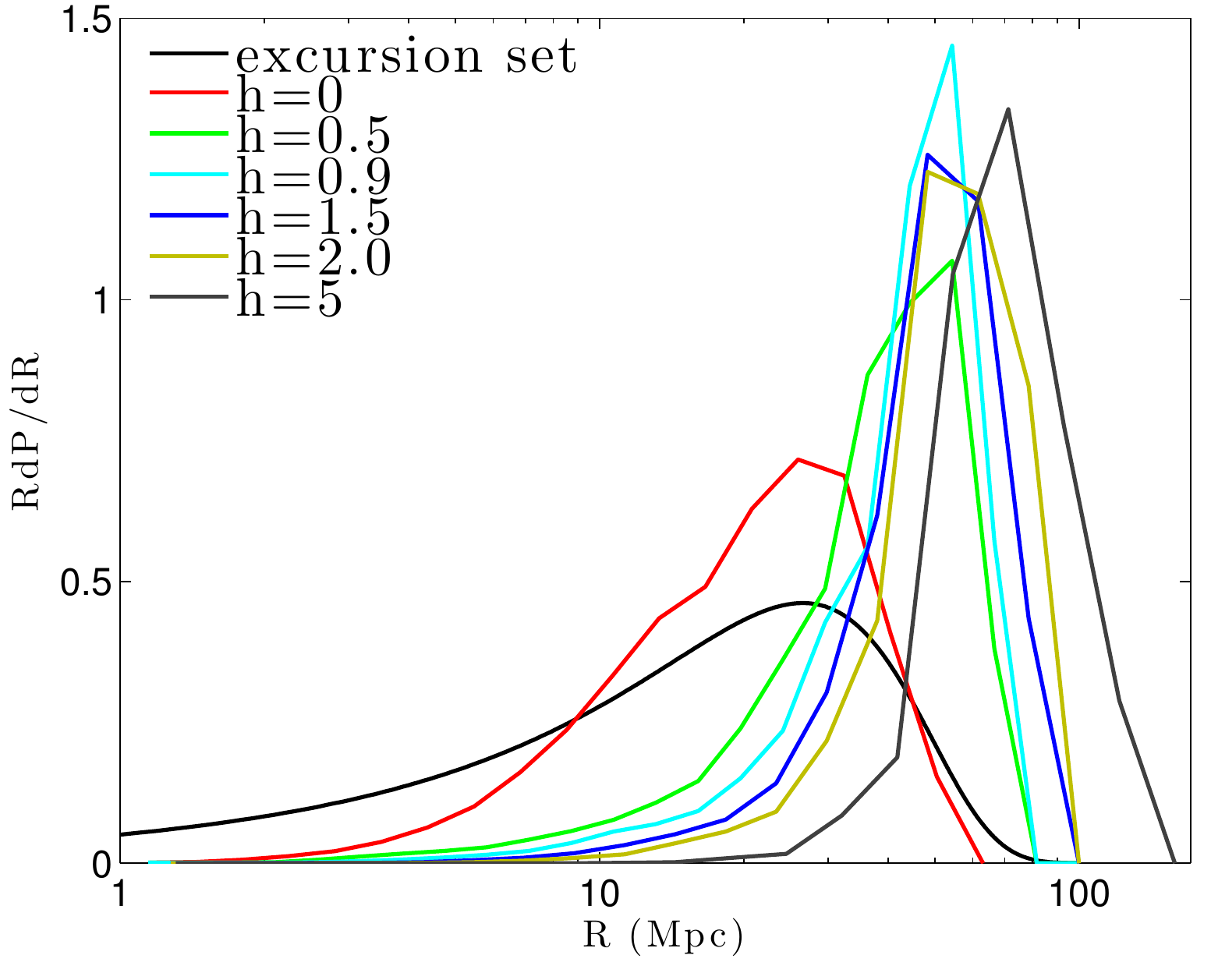}

    \caption{Bubble PDF derived by the watershed algorithm for the Monte-Carlo simulation (top) and semi-numeric simulation boxes at $Q_{\text{HII}} =0.28,0.49,0.78$ ($z=12,11,10$; second to bottom) for varying smoothing parameters $h$.}
    
\label{fig:monte_mass_function_vary_h}
\end{center}
\end{figure}

\begin{figure}
\begin{center}
\includegraphics[scale=0.35]{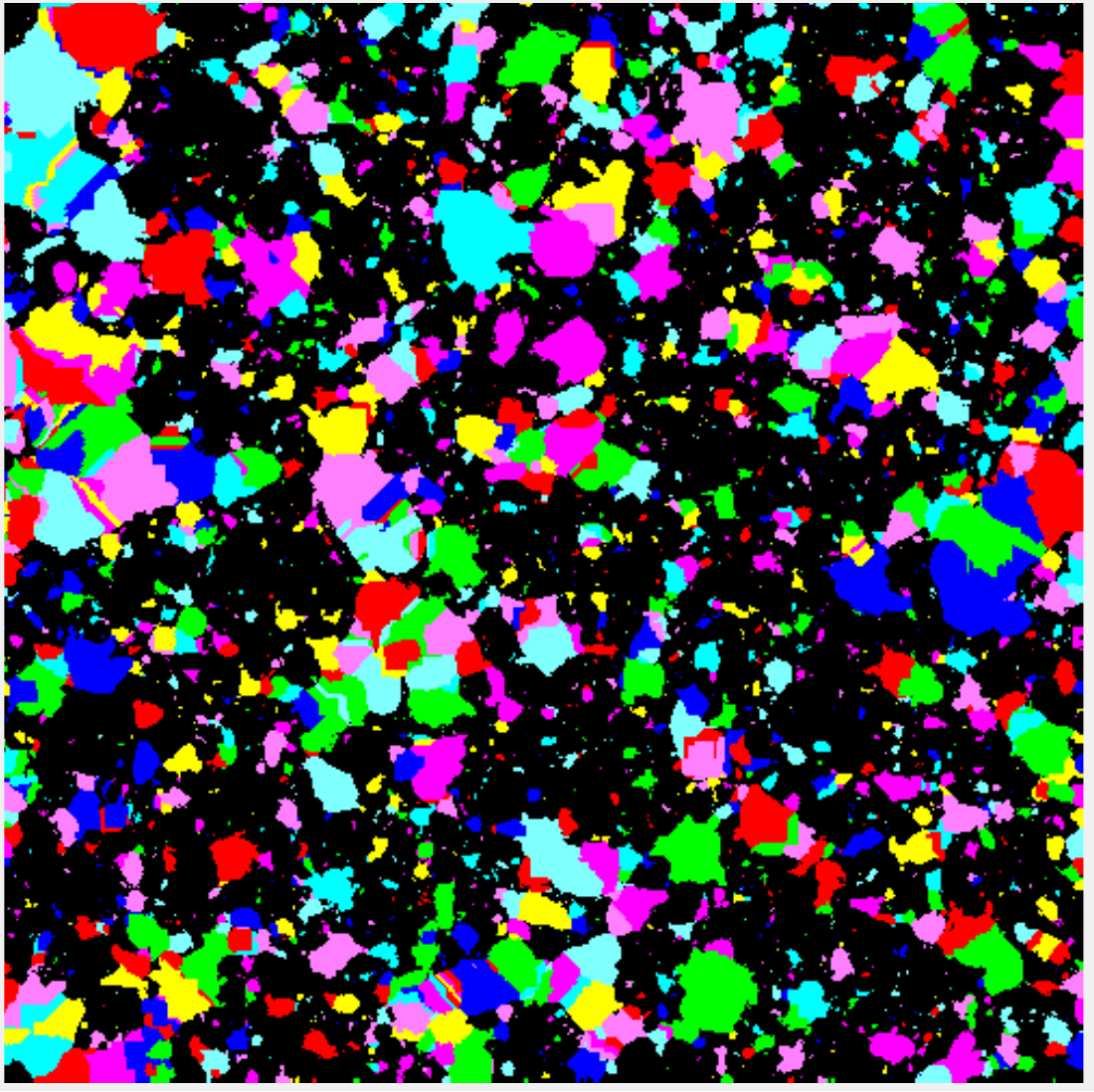}
\includegraphics[scale=0.35]{./slice_z11_h09_500Mpc}
\includegraphics[scale=0.35]{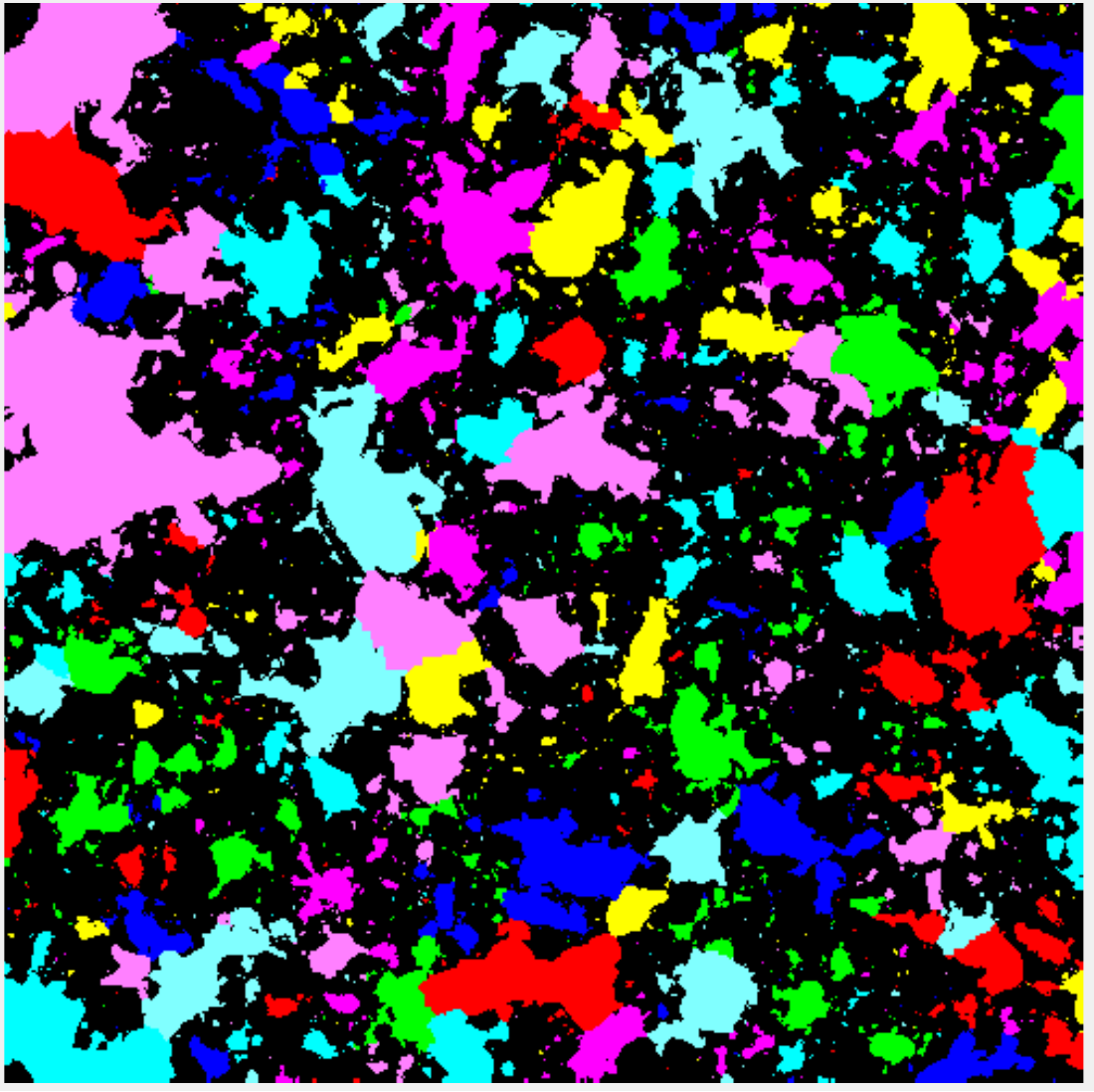}
\includegraphics[scale=0.35]{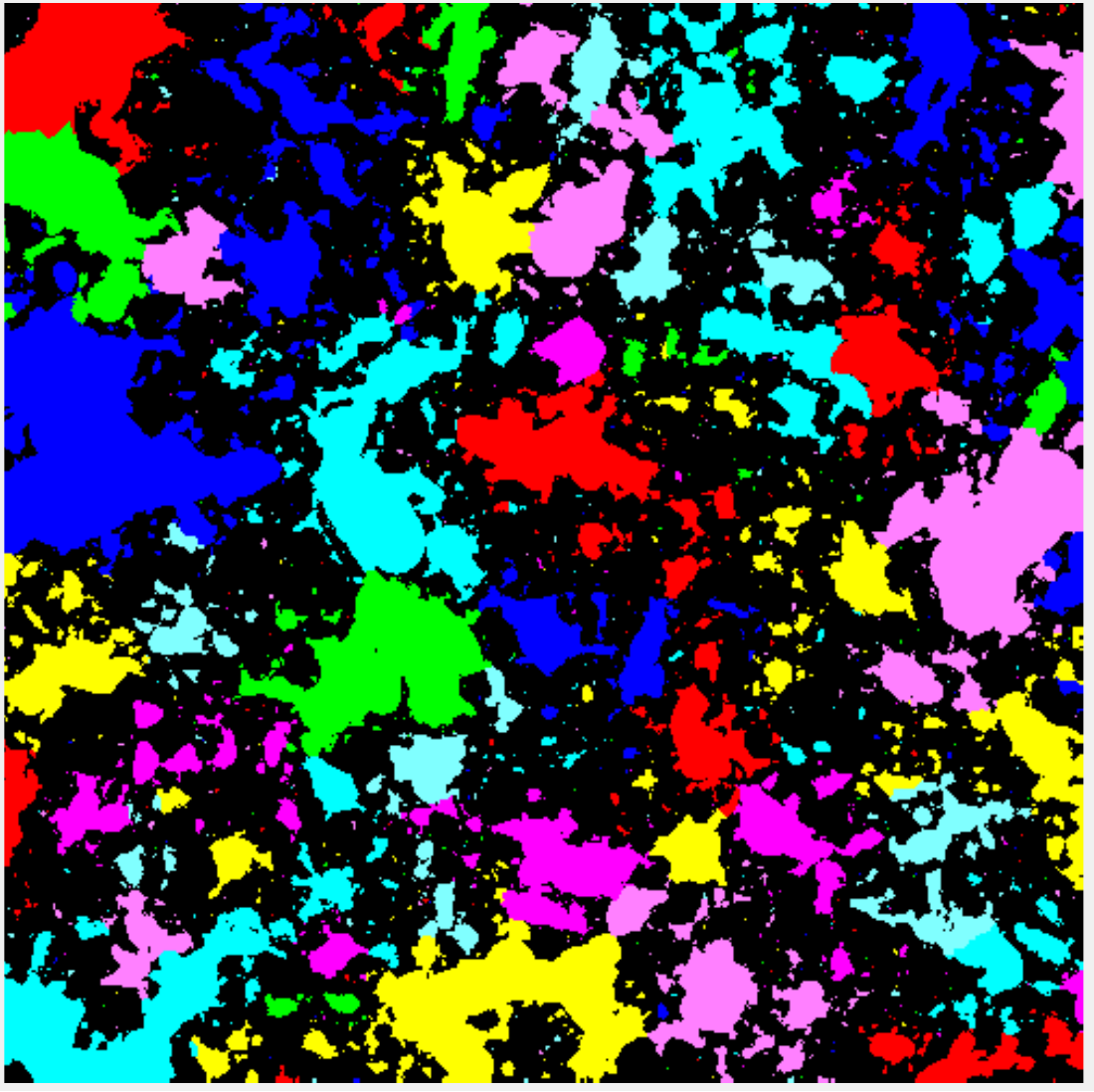}
    \caption{Clockwise from top left are watershed segmented slices of real simulation box with $Q_{\text{HII}} =0.49$ ($z=11$), with smoothing parameters of $h=0, 0.9, 2, 5$. The oversegmentation is significant for $h=0$. Once we tune up the values of $h$, the results are more realistic segmented bubbles as seen in slices with $h=0.9$, $2$ and $5$.}
    \label{fig:slices_smoothing}
\end{center}
\end{figure}

Now let us examine the effects of the smoothing parameter. The top panel of Fig \ref{fig:monte_mass_function_vary_h} shows the bubble size distribution for various choices of $h$, for the unclustered bubbles simulation.  Over the range $h=0-2$, the recovered PDF is robust to the choice of $h$, and in good agreement with the underlying PDF. For higher values of $h$, the bubble size distribution is artificially truncated (e.g, the $h=5$ PDF is truncated at $R=1$ Mpc, as appropriate for 0.2 Mpc voxels), artificially biasing the bubble PDF toward larger sizes. Clearly, the algorithm is performing as one might expect. However, this case is not particularly informative: because of the lack of clustering, there is less bubble overlap, fewer local minima, and hence less need for smoothing (even $h=0$ returns sensible results). The next 3 panels show how the bubble PDFs changes as a function of smoothing parameter for realistic semi-numeric simulation boxes, for ionization fractions $Q_{\rm HII}=0.28,0.49,0.78$ ($z=12,11,10$). While there is more sensitivity to smoothing parameter (likely related to ambiguities in segmenting the large bubbles which arise from clustering), there is nonetheless an `intermediate asymptotic' range where the bubble PDF depends only weakly on smoothing. We have employed smoothing in this range in all the subsequent plots we presented. Visually, we can see the effect of smoothing in Fig \ref{fig:slices_smoothing}, which presents slices of the watershed segmented box for different smoothing parameters. Without smoothing, the bubbles appear significantly over-segmented. As the smoothing scale $h$ increases, there is a range where segmentation is fairly stable and appears visually sensible.

Ultimately, the fact that there is an adjustable smoothing parameter introduces an unavoidable degree of freedom into the watershed algorithm. The bubble sizes tend to increase as the smoothing parameter $h$ increases. For no smoothing ($h\rightarrow 0$), all local minima are identified as bubble centers, and for structures where there is a good deal of overlap between bubbles (due to clustering; this is certainly true of reionization), the watershed algorithm gives a mass function similar to that of a straight distance transform. At the other end of the spectrum, for larger amounts of smoothing (as h approaches the box size), the watershed identifies only collection basins with very deep minima, and becomes equivalent to the friends-of-friends algorithm. For realistic simulations of reionization, this results in most of the ionized volume residing in a gigantic bubble spanning the box. We argue that sensible choices for the smoothing parameter $h$ can be identified, based on two criteria: (i) an 'intermediate asymptotic' range where the bubble PDF is independent or only very weakly dependent on $h$; (ii) reasonable agreement in peak bubble size with the MFP algorithm. The latter is a physically well-motivated algorithm which shows good agreement with watershed in the well-understood case where bubble overlap is insignificant. Finally, a visual check of watershed segmentation results is always helpful. 

Of course, if one is applying the watershed algorithm to observations with finite resolution, the choice of smoothing parameter is obvious-- indeed, in this case the smoothing has already been done. The observed PDF is heavily distorted by the broad window function of the observations. We leave analysis of this situation for future work. 

\section{Discussion}
\label{section:discussion} 

As we have seen, there is no unique definition of bubble size, and different methods will highlight different aspects of the complicated topology of HII regions. Nonetheless, it is worth considering the most appropriate methods for questions we are likely to be asking. There are two compelling observational motivations. One is the hope of  performing 21cm tomography with next generation instruments such as HERA\footnote{http://reionization.org}. One would like to know the angular resolution needed at different stages of reionization to actually image the HII regions\footnote{Frequency resolution -- i.e., resolution in the line of sight direction -- is generally much higher and not a limiting factor.}. This corresponds to the typical size of features in 2D slices of the simulation box. It is a task which the watershed algorithm is ideally suited to perform, given its origins as a well-tested, canonical image segmentation algorithm in 2D. From Fig \ref{fig:scales}, we see that the watershed algorithm does provide a reasonable measure of typical bubble sizes. It is also equally obvious that the excursion set formalism predicts bubble sizes that are too small. Another important observational application is the mean free path of ionizing photons, which determines the amplitude of the ionizing background $\Gamma$ at these high redshifts (via $\Gamma \propto \epsilon \lambda$, where $\epsilon$ is the comoving emissivity and $\lambda$ is the mean free path). This is closely related to the mean free path of Ly$\alpha$ photons, which determines if Ly$\alpha$ photons can redshift out of resonance in ionized bubbles before being scattered by neutral hydrogen. Ly$\alpha$ emitters will be visible in large bubbles, but not in smaller ones where this condition is not satisfied. The visibility and clustering of Ly$\alpha$ emitters, and the redshift evolution of these properties, has been proposed as a sensitive diagnostic of reionization \citep{mcquinn07-lya,treu13}. Both these applications are obviously closely related to the MFP statistic\footnote{The correspondence is not exact, because sources tend to be biased toward the center of bubbles rather than occupy random positions. In principle, it should be possible to do a more careful job, given known positions of halos in semi-numeric simulations, but we leave this for future work.}. Thus, at present we regard the watershed and MFP methods as the best means for estimating observationally relevant bubble sizes. By contrast, the DT (which underestimates bubble sizes) and FoF methods (which overestimates bubble sizes, as it focusses solely on connectivity) return answers which at present appear less relevant to observations. The fact that both the watershed and MFP bubble size distributions peak at approximately the same physical scales gives us additional confidence. 

In considering the characteristic scale of bubbles, there is also the broader question of how to partition ionized regions, and when bubbles should be regarded as physically distinct. This is a question which only the watershed and FoF algorithms can address, as only they are capable of segmentation. Consider two large spherical bubbles connected by a thin tube. Instinctively we would decompose this into 3 components (the two bubbles and the tube), and indeed this is what the watershed algorithm returns. By contrast, the FoF algorithm would characterize this as a single ionized region. Indeed, the FoF or any related percolation algorithm (such as the Hoshen-Kopelmann algorithm) where the linking length is allowed to be arbitrarily large, determines that most of the volume is in a single, large ionized region for $x_{\rm HII} > 10\%$ \citep{furlanetto16}. At this point, when almost all ionized points are connected but their filling factor is small, the HII regions have a 'tunnel-like' topology. The connectivity of a network of tunnels is obviously important for transport processes which obey mass conservation--e.g., coffee percolating though coffee grounds in a filter, groundwater percolating through holes in the earth, or a Pac-man trying to traverse a maze. Its importance is less obvious for reionization, where (modulo the effects of recombinations to the ground state) ionizing photons travel in straight lines and are absorbed once they hit the HII region boundary. The photons cannot follow bends in the tunnel.\footnote{Nonetheless, FoF yields important physical insights. For instance, all the bubbles in Fig \ref{fig:scales} actually connect to one another, so receive some illumination from sources at much larger scales than the watershed algorithm would suggest (particularly since recombinations to the ground state allow for some scattering of ionizing photons). The fact that during the middle stages of reionization, there is an infinite ionized region intertwined with an infinite neutral region is a physical fact which is not apparent in 2D slices or from other algorithms \citep{furlanetto16}. It is a robust prediction which potentially serves as a stringent test of future high-resolution interferometers. In this case, what is perhaps more relevant is the size of a typical cross-section --i.e. the width of these curved tubes, which dominate the solid-angle weighted\footnote{This solid angle weighting is taken into account in our calculation of the mfp window function in Appendix A.}} photon mean free paths. From the MFP algorithm (which also concurs with the watershed algorithm), we conclude that these {\it do} have a characteristic scale. The existence of characteristic scales in 2D slices is why the excursion set theory predictions of a typical size scale was for so long uncritically accepted. Even though we have shown that the quantitative predictions of excursion set theory need to be revised, we still maintain that the notion of a characteristic bubble size is correct and physically meaningful. 

Let us consider our example of 2 spherical bubbles, and bring them closer together. At the point at which they just begin to overlap, by the lights of the observational applications we have mentioned (e.g., photon mean free path), it still makes sense to regard them as distinct bubbles. However, as the level of overlap increases, as some point the bubbles are best described as having merged, constituting a single entity. The dividing line between these two cases can be somewhat indistinct, and obviously application and geometry dependent. Nonetheless, the watershed algorithm provides an objective, well-defined way of performing this segmentation, which has been extensively tested in the image processing community. As we have seen, in most cases it agrees well with what we would visually classify as distinct regions. 

In this paper, we have argued that excursion set theory underpredicts bubble sizes, particularly early on in reionization. There are two questions worth considering in this regard: (i) what causes FZH04 to fail? (ii) if excursion set theory as formulated in FZH04 is defective, why is it nonetheless acceptable to use semi-numeric simulations? Both are after all based on the excursion set theory approach --- smoothing the halo density field and marking the largest possible spherical volumes which are able to self-ionize.

With regards to point (i), the close correspondence at all stages of reionization between bubble sizes predicted by FZH04 and uncovered from simulations by the DT method (Fig \ref{fig:dist_vs_unconv}) is noteworthy. In particular, the DT method infers a rapid evolution in bubble size which the other two methods do not. This may not be accidental. The DT method is very closely allied to the excursion set theory approach. At each point in the simulation box, the DT method smooths over the ionization field (rather than the density field) to find the largest spherical region which can be accommodated within an HII region, and labels that as a bubble of radius R. This raw distance transform overweights voxels at the bubble boundary, leading to an underestimate of bubble size. This close correspondence meant that the DT method was mistakenly used to {\it confirm} the results of analytic excursion set theory from simulations. In reality, it might give very similar results because it is a very similar method by construction, and suffers from the same defects. The effect is particularly egregious for the tube-like structures (caused by merger of similar-sized bubbles), which appear early in the history of reionization. These bubbles have large inverse porosity, i.e. they subtend large stretches of space without filling them. The over-counting of small bubbles by the DT method (and similarly by excursion set theory) is obvious if one thinks about the contours of constant distance transform. By contrast, the watershed algorithm adds a secondary processing step (flooding) which joins all connected regions with the same distance transform, thereby avoiding this bias. For instance, a straight tube will be identified as a single HII region. The mean free path algorithm also gives an acceptably unbiased answer since all trajectories are weighted by solid angle. Interestingly, FZH04 and the distance transform both give answers closer to the watershed algorithm at higher ionization fraction, when the ratio of surface area to volume is smaller, and boundary effects are less important. Bubbles become increasingly quasi-spherical as reionization proceeds \citep{furlanetto16}, and the explicit assumption of sphericity is approximately correct. 

Alternatively, a modified excursion set theory may be capable of producing the correct bubble distribution. A particularly promising approach is the model of \citet{paranjape14}, which predicted qualitative changes to the FZH04 bubble PDF -- bubble sizes larger by a factor of 2-- which are in agreement with the results of this paper. It proposes two modifications to standard excursion set theory: (i) standard excursion set theory assumes that random walks are uncorrelated, even though neighboring walks sample the same large scale fluctuations and real-space filtering (used to find the ionization pattern) couples different Fourier modes. This effect can now be accounted for in analytic models \citep{musso12,musso14}. (ii) One should not consider spherical averages (and random walks) at all locations, but only at special ones. Both the most massive halos and the largest cosmic voids form at the most extreme peaks of the density field, invalidating the usual excursion set assumption that (e.g.) halos enclose a single density contrast. \citet{paranjape14} find that the peaks constraints on halo locations has the largest effect. While their models and our simulation results do not fully agree in detail, these promising insights clearly deserve more work. We plan more detailed comparisons in the future.

Although the FZH04 excursion set theory based approach is incorrect for counting bubbles, as seen by its failure to correctly predict the mass function, excursion set theory is nonetheless correct for painting on the ionization field. This can be seen from the close agreement between semi-numeric simulations utilizing this approach with full radiative transfer simulations \citep{mcquinn07,zahn07,zahn11}. This is because the basic photon-counting argument behind the excursion set theory approach, which predicts the location of HII region boundaries--assuming they travel in straight lines--is correct. Difficulties only arise when one uses this approach directly to count bubbles. While the straight line assumption works well locally, it cannot fully characterize the stochastic overlap of realistic HII regions. As we have argued, the partitioning of a contiguous ionized region into separate bubbles is non-trivial. The watershed approach handles this by identifying local minima, which are effectively the bubble centers. The distance from these points to HII region boundaries is more accurate than unrefined distance transforms.

Regardless of the scheme we use, it is clear that bubble sizes are consistently underestimated in excursion set theory, by a factor $\sim 2-10$ in effective radius (so a factor $\sim 10-1000$ in volume), with the largest disparity during the early stages of reionization. Bubble sizes in the simulations clearly evolve more weakly with redshift/ionization filling fraction than in excursion set theory, an effect already noticed by \citet{mesinger07}. This arises because mergers of neighboring bubbles happens very early on, an effect which can be understood from percolation theory \citep{furlanetto16}. In percolation theory, an object of infinite extent (in practice, an object which spans the box) arises at a certain critical filling fraction, i.e., at that point, the structure percolates the entire volume. If the bubbles are Poisson distributed, this fraction is $\sim 30\%$ \citep{stauffer94}. However, for a Gaussian random field this critical fraction drops due to clustering; for $\Lambda$CDM parameters the critical fraction is $\sim 10\%$ \citep{shandarin06,furlanetto16}. The clustering and percolation of bubbles is apparently not correctly handled by the analytic expressions of FZH04.

\section{Conclusions}
\label{section:conclusions} 

Our conclusions are as follows: 
\begin{itemize}
\item{The watershed algorithm is a superior method for segmenting bubbles and inferring its mass function. It has the narrowest window function for spherical bubbles (a delta function), and is unbiased. It provides the best correspondence to intuitive visual segmentation of images, and has the best performance on controlled Monte-Carlo samples. It does have an adjustable smoothing parameter, but we have found robust results over a reasonable range of this parameter.}

\item{The mean free path method is another good complementary method, and corresponds to an important physical quantity. By contrast, the distance transform (equivalent to spherical averaging used by previous authors) is unacceptably biased. The friends of friends algorithm, while focusing on a different aspect of bubble topology, does yield important physical insights, and it is imperative to translate these physical insights into observationally testable predictions.}

\item{Both the watershed and MFP methods still show that there is a characteristic bubble size which increases with time during reionization. However, these bubble sizes are significantly larger (by up to an order of magnitude), and evolve more slowly with ionization fraction, than predicted by excursion set theory. The largest disparity occurs early in reionization. In addition, the bubble size distribution is narrower than predicted, and becomes increasingly sharply peaked as reionization progresses. Excursion set theory apparently does not correctly handle non-spherical structures produced by bubble mergers, particularly when the ionization filling factor is small, and should not be used to predict characteristic scales.}

In the future, it would be interesting to examine the size distribution of neutral regions with the watershed algorithm. This is an important, observationally accessible quantity during the late stages of reionization. However, semi-numeric simulations become unreliable toward the tail end of reionization, when photon mean free paths are limited by LLSs (which can only be incorporated in an ad-hoc fashion) rather than bubble sizes. Such a study requires analysis of simulations with full radiative transfer, including Lyman limit systems (LLSs); the latter can become prohibitively expensive, due to the wide required dynamic range ($\Delta x \lsim 0.1$ kpc resolution to accurately resolve LLSs, to the $L \gsim 300$ Mpc box sizes required to reduce cosmic variance for the largest volume-filling bubbles). It is possible also to consider many refinements: exploring the many variants of watershed or other segmentation/tesselation algorithms, refining the mean free path technique (e.g., using maximum likelihood to infer the underlying bubble distribution given a known window function; casting rays only from ionizing halo locations), or exploring other methods. In particular, it would be very interesting to consider whether the watershed algorithm can be applied directly to observations. Another obvious line of attack would be to arrive at a physical understanding of the characteristic scales and bubble PDFs we have obtained. The parallels between cosmic voids and HII bubbles in reionization, where many of the same techniques have been used (excursion set theory, segmentation algorithms) may also be a source of further insight. 

\end{itemize}

\section*{Acknowledgments}

This work was completed as part of the University of California Cosmic Dawn Initiative. We acknowledges support from the University of California Office of the President Multicampus Research Programs and Initiatives through award MR-15-328388. YL acknowledges a UCSB College of Creative Studies SURF fellowship. SPO acknowledges NASA grants NNX12AG73G and NNX15AK81G. SRF was partially supported by a Simons Fellowship in Theoretical Physics. SRF also thanks the Observatories of the Carnegie Institute of Washington for hospitality while much of this work was completed. We thank Andrei Mesinger for helpful discussions. 

\begin{appendix} 
\section{Derivation of MFP Window Function} 
\label{appendix:mfp_window_func} 

In this Appendix, we will derive the window function $W(r/R)$ for the mean free path (MFP) method (\S \ref{section:mfp}). This is equivalent to the conditional probability density function $P(r|R)$ for the mean free path of distance $r$ inside a sphere of radius $R$. Many of the steps here follow the derivation as presented in \citet{solomon78}, with some modifications. 

\begin{figure}
\includegraphics[width=0.5\textwidth]{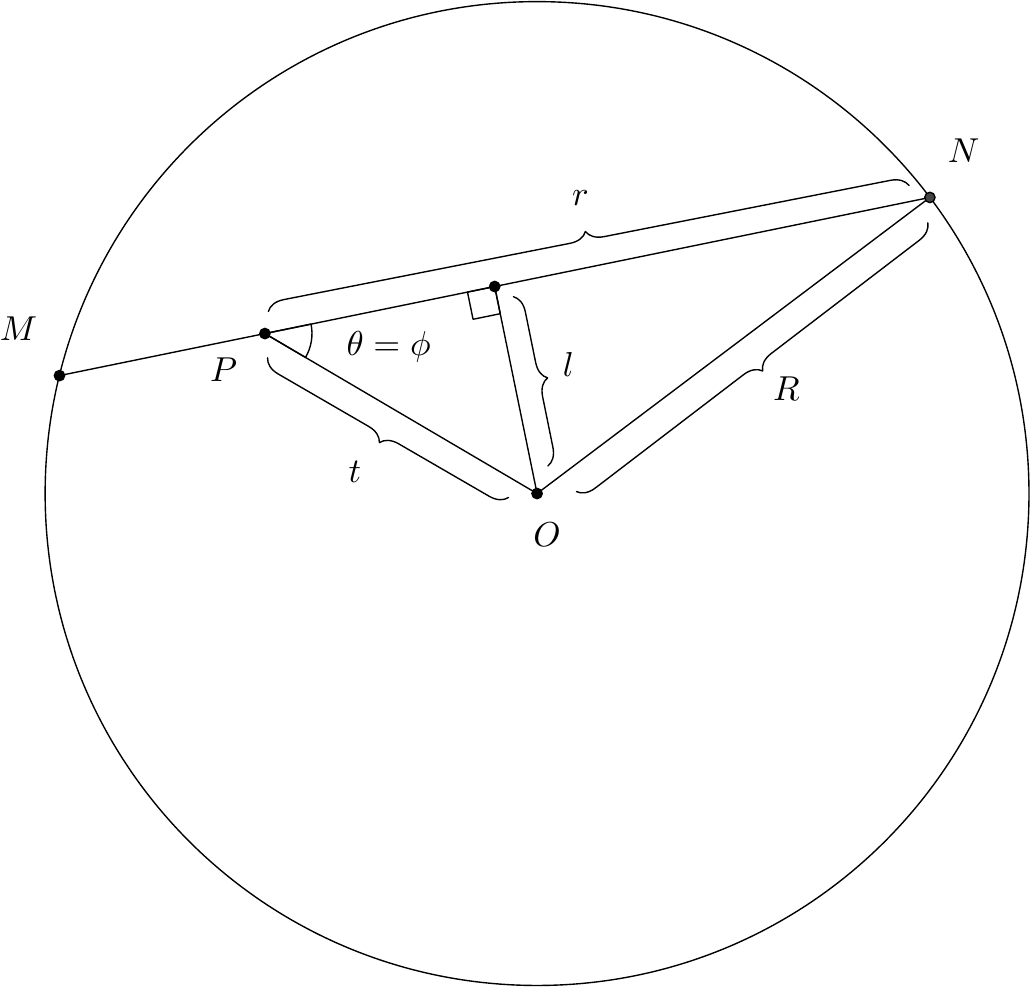} \\
\includegraphics[width=0.5\textwidth]{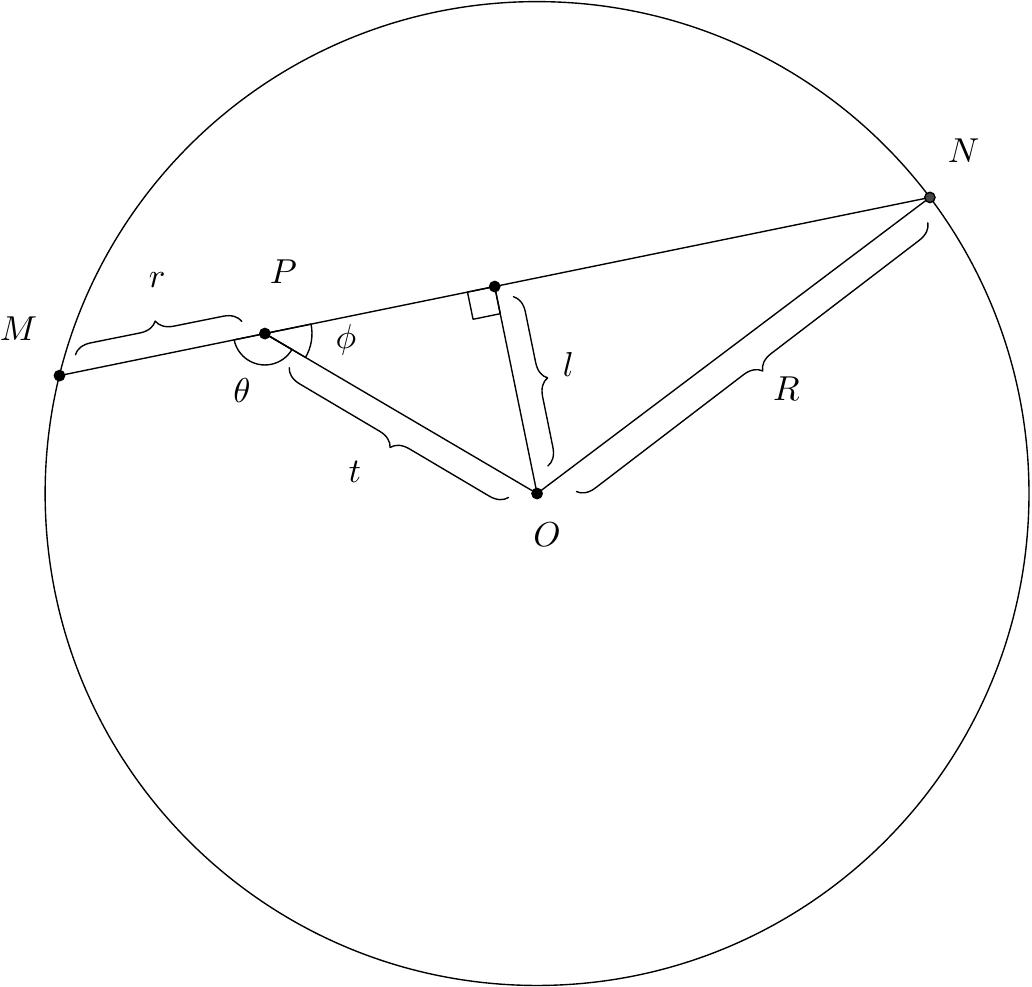}
\caption{Geometry for calculation of window function for mean free path (MFP) method. Top panel: acute $\theta$. Bottom panel: obtuse $\theta$.}
\label{fig:mfp_geometry}
\end{figure}

There are two steps in our construction: first we choose a random point, $P$, inside the sphere, then we choose a random direction to project the line and record its length $r$. Let $t$ be the distance between the center and the random point P. The density function of $t$ is proportional to the volume of the spherical shell, $g(t) dt = 4 \pi t^{2} dt /(4 \pi/3 R^{3})$, or:
\begin{equation}
g(t) = \frac{3t^2}{R^3}~~(0\leq t \leq R)
\end{equation}
Let $\theta$ be the angle between the line through the origin $O$ and the mean free path cord. The density function is given by
\begin{equation}
h(\theta) = \frac{\sin(\theta)}{2} ~~ \left(0\leq \theta \leq \pi\right)
\end{equation} 
Let $\phi$ be the acute angle between the line through the origin $O$ and the mean free path cord $MN$, so for acute $\theta$, $\theta= \phi$; for obtuse $\theta$, $\theta = \pi - \phi$ (see Fig \ref{fig:mfp_geometry}). Let $l$ be the length of perpendicular line from $O$ to our mean free path chord. We know $ l = t\sin(\phi) = t\sin(\theta)$ and $dl = t\cos(\theta)d\theta$ for a fixed t. Then the conditional density of $l$ for a fixed $t$ is 
\begin{equation}
h(l|t) = \sin(\theta)\left|\frac{1}{2t\cos(\theta)}\right| = \left | \frac{\tan(\theta)}{2t} \right|
~~(0\leq \theta \leq \pi)
\end{equation}
where the absolute sign eliminates negative probability density for obtuse $\theta$. This can be written in terms of $l$, $t$, and $r$ as
\begin{equation}
h(l|t)  = \left| \frac{\tan(\theta)}{2t} \right|  = \left|\frac{\tan(\phi)}{2t}\right| = \pm \frac{l}{2t\left(r - \sqrt{R^2 - l^2}\right)}
\end{equation}
where plus/minus is for acute/obtuse $\theta$ respectively.

The conditional probability density function $P(l|R)$ is the product of $h(l|t)$ and $g(t)$ integrated over all possible $t$:
\begin{equation}
P(l|R) =\int_l^R h(l|t)g(t) dt = \pm\int_l^R \left(\frac{3tl}{2R^3(r - \sqrt{R^2-l^2})}\right) dt
\end{equation}
Now we want to perform a change of variables to express $l$ in terms of $r$ and $t$ to find density $P(r|R)$. We know that 
$\sqrt{t^2 - l^2} \pm \sqrt{l^2 + R^2} = r$ where plus/minus is for acute/obtuse $\theta$. Solving this equation for $l$ and discarding extraneous solutions we obtain: 
\begin{equation}
l = \frac{L^{2}}{2r}
\label{eqn:l}
\end{equation}
where
\begin{equation} 
L^{2}=\sqrt{-r^4+2r^2R^2+2r^2t^2-R^4+2R^2t^2-t^4}
\end{equation} 
and
\begin{align}
\frac{\partial l}{\partial r} &= \frac{-r^2 + R^2 + t^2}{L^{2}} -
\frac{L^{2}}{2r^2}\\
&= \frac{(R^2 - t^2 + r^2)(R^2 - t^2 - r^2)}{2r^2L^{2}}
\label{eqn:dldr}
\end{align}
so the absolute value of the Jacobian determinant is:
\begin{equation}
|J| = \left|\frac{\partial l}{\partial r}\right| = \mp\frac{\partial l}{\partial r}
\end{equation}
where again minus/plus is for acute/obtuse $\theta$. This combines with the previous plus and minus sign to give an overall minus sign.

Lastly, we need to figure out the limit of integration for $t$ for a fixed $r$. The upper limit is $R$ as it cannot exceed the radius of the sphere. However, the lower limit for  $t$ is $\left| R-r \right|$ which occurs when $\theta = 0$. Now plug in everything in hand and we have: 
\begin{align}
P(r|R)= \int ^R_{\left| R-r \right|} dt 
&\left[\frac{3tL^{2}}{2 r R^3}\right] \times \nonumber \\
&\left[r - \sqrt{R^2-\left((2r)^{-1}L^{2}\right)^2}\right]^{-1} \times\nonumber\\
&\left[-\frac{(R^2 - t^2 + r^2)(R^2 - t^2 - r^2)}{2r^2L^{2}}\right]\nonumber\\
\label{eqn:fl}
\end{align}
This analytic result matches numerical Monte-Carlo results. 

Note that this result differs from the erroneous expression given in \citet{mesinger07}. The latter can be derived by calculating the probability of choosing two random points, one on the interior of the sphere and one on its surface, and calculating the distance between them. However, this approach does not correctly weight by solid angle. For instance, a random ray emerging from a point close to the bubble surface is much more likely to have a small MFP, due to the enhanced likelihood of striking the nearby surface. The \citet{mesinger07} window function, which fails to take this into account, is skewed towards larger bubble sizes. When this incorrect window function is convolved with analytic predictions, it boosts typical bubble sizes and reduces the discrepancy between analytic predictions and simulation results.

\end{appendix}

\bibliography{bubble_paper_final}

\begin{thebibliography}{26}
\expandafter\ifx\csname natexlab\endcsname\relax\def\natexlab#1{#1}\fi

\bibitem[{{Finlator} {et~al}\mbox{.}(2012){Finlator}, {Oh}, {{\"O}zel}, \&
  {Dav{\'e}}}]{finlator12}
{Finlator} K., {Oh} S.~P., {{\"O}zel} F., {Dav{\'e}} R., 2012, \mnras, 427,
  2464

\bibitem[{{Friedrich} {et~al}\mbox{.}(2011){Friedrich}, {Mellema}, {Alvarez},
  {Shapiro}, \& {Iliev}}]{friedrich11}
{Friedrich} M.~M., {Mellema} G., {Alvarez} M.~A., {Shapiro} P.~R., {Iliev}
  I.~T., 2011, \mnras, 413, 1353

\bibitem[{{Furlanetto} \& {Oh}(2005)}]{furl05-rec}
{Furlanetto} S.~R., {Oh} S.~P., 2005, \mnras, 363, 1031

\bibitem[{{Furlanetto} \& {Oh}(2016)}]{furlanetto16}
{Furlanetto} S.~R., {Oh} S.~P., 2016, \mnras, 457, 1813

\bibitem[{{Furlanetto} {et~al}\mbox{.}(2004){Furlanetto}, {Zaldarriaga}, \&
  {Hernquist}}]{furl04-bub}
{Furlanetto} S.~R., {Zaldarriaga} M., {Hernquist} L., 2004, \apj, 613, 1

\bibitem[{{Iliev} {et~al}\mbox{.}(2006){Iliev}, {Mellema}, {Pen}, {Merz},
  {Shapiro}, \& {Alvarez}}]{iliev06}
{Iliev} I.~T., {Mellema} G., {Pen} U.-L., {Merz} H., {Shapiro} P.~R., {Alvarez}
  M.~A., 2006, \mnras, 369, 1625

\bibitem[{{McQuinn} {et~al}\mbox{.}(2007{\natexlab{a}}){McQuinn}, {Hernquist},
  {Zaldarriaga}, \& {Dutta}}]{mcquinn07-lya}
{McQuinn} M., {Hernquist} L., {Zaldarriaga} M., {Dutta} S., 2007{\natexlab{a}},
  \mnras, 381, 75

\bibitem[{{McQuinn} {et~al}\mbox{.}(2007{\natexlab{b}}){McQuinn}, {Lidz},
  {Zahn}, {Dutta}, {Hernquist}, \& {Zaldarriaga}}]{mcquinn07}
{McQuinn} M., {Lidz} A., {Zahn} O., {Dutta} S., {Hernquist} L., {Zaldarriaga}
  M., 2007{\natexlab{b}}, \mnras, 377, 1043

\bibitem[{{Mesinger} \& {Furlanetto}(2007)}]{mesinger07}
{Mesinger} A., {Furlanetto} S., 2007, \apj, 669, 663

\bibitem[{{Musso} \& {Sheth}(2012)}]{musso12}
{Musso} M., {Sheth} R.~K., 2012, \mnras, 423, L102

\bibitem[{{Musso} \& {Sheth}(2014)}]{musso14}
{Musso} M., {Sheth} R.~K., 2014, \mnras, 443, 1601

\bibitem[{{Neyrinck}(2008)}]{neyrinck08}
{Neyrinck} M.~C., 2008, \mnras, 386, 2101

\bibitem[{{Paranjape} \& {Choudhury}(2014)}]{paranjape14}
{Paranjape} A., {Choudhury} T.~R., 2014, \mnras, 442, 1470

\bibitem[{{Paranjape} {et~al}\mbox{.}(2016){Paranjape}, {Choudhury}, \&
  {Padmanabhan}}]{paranjape16}
{Paranjape} A., {Choudhury} T.~R., {Padmanabhan} H., 2016, \mnras

\bibitem[{{Petrovic} \& {Oh}(2011)}]{petrovic11}
{Petrovic} N., {Oh} S.~P., 2011, \mnras, 413, 2103

\bibitem[{{Planck Collaboration} {et~al}\mbox{.}(2015){Planck Collaboration},
  {Ade}, {Aghanim}, {Arnaud}, {Ashdown}, {Aumont}, {Baccigalupi}, {Banday},
  {Barreiro}, {Bartlett}, \& et~al.}]{planck-collaboration15}
{Planck Collaboration} {et~al.}, 2015, ArXiv e-prints

\bibitem[{{Platen} {et~al}\mbox{.}(2007){Platen}, {van de Weygaert}, \&
  {Jones}}]{platen07}
{Platen} E., {van de Weygaert} R., {Jones} B.~J.~T., 2007, \mnras, 380, 551

\bibitem[{{Shandarin} {et~al}\mbox{.}(2006){Shandarin}, {Feldman}, {Heitmann},
  \& {Habib}}]{shandarin06}
{Shandarin} S., {Feldman} H.~A., {Heitmann} K., {Habib} S., 2006, \mnras, 367,
  1629

\bibitem[{{Sobacchi} \& {Mesinger}(2014)}]{sobacchi14}
{Sobacchi} E., {Mesinger} A., 2014, \mnras, 440, 1662

\bibitem[{Soille(2013)}]{soille13}
Soille P., 2013, Morphological image analysis: principles and applications.
  Springer Science \& Business Media

\bibitem[{Solomon(1978)}]{solomon78}
Solomon H., 1978, Geometric probability, CBMS-NSF regional conference series in
  applied mathematics. Society for Industrial and Applied Mathematics,
  Philadelphia

\bibitem[{Stauffer \& Aharony(1994)}]{stauffer94}
Stauffer D., Aharony A., 1994, Introduction to percolation theory. CRC press

\bibitem[{{Sutter} {et~al}\mbox{.}(2015){Sutter}, {Lavaux}, {Hamaus}, {Pisani},
  {Wandelt}, {Warren}, {Villaescusa-Navarro}, {Zivick}, {Mao}, \&
  {Thompson}}]{sutter15}
{Sutter} P.~M. {et~al.}, 2015, Astronomy and Computing, 9, 1

\bibitem[{{Treu} {et~al}\mbox{.}(2013){Treu}, {Schmidt}, {Trenti}, {Bradley},
  \& {Stiavelli}}]{treu13}
{Treu} T., {Schmidt} K.~B., {Trenti} M., {Bradley} L.~D., {Stiavelli} M., 2013,
  \apjl, 775, L29

\bibitem[{{Zahn} {et~al}\mbox{.}(2007){Zahn}, {Lidz}, {McQuinn}, {Dutta},
  {Hernquist}, {Zaldarriaga}, \& {Furlanetto}}]{zahn07}
{Zahn} O., {Lidz} A., {McQuinn} M., {Dutta} S., {Hernquist} L., {Zaldarriaga}
  M., {Furlanetto} S.~R., 2007, \apj, 654, 12

\bibitem[{{Zahn} {et~al}\mbox{.}(2011){Zahn}, {Mesinger}, {McQuinn}, {Trac},
  {Cen}, \& {Hernquist}}]{zahn11}
{Zahn} O., {Mesinger} A., {McQuinn} M., {Trac} H., {Cen} R., {Hernquist} L.~E.,
  2011, \mnras, 414, 727

\end{thebibliography}

\end{document}